\documentclass[pra,aps,twocolumn,nopacs,superscriptaddress,nofootinbib]{revtex4}

\usepackage{graphicx}  
\usepackage{dcolumn}  
\usepackage{bm}           
\usepackage{bbm}
\usepackage{amsmath}
\usepackage{amssymb}
\usepackage{amsfonts}
\usepackage{enumerate}
\usepackage{indentfirst}
\usepackage{psfrag}
\usepackage{float}
\usepackage{color}

\def\ii{{\rm i}}  \def\ee{{\rm e}}

\def\rb{{\bf r}}  \def\Rb{{\bf R}}

  \def\kpar{k_\parallel}  \def\kparb{{\bf k}_\parallel}
\def\Qb{{\bf Q}}    
\def\me{m_{\rm e}}  

\def\vF{v_{\rm F}}  \def\kF{{k_{\rm F}}}  \def\EF{{E_{\rm F}}}
        \def\wp{\omega_{\rm p}}
  
  \def\epsb{\eps_{\rm b}}
\def\eps{\epsilon}  \def\dm{d_{\rm m}}
    \def\en{\varepsilon}

\begin{document}
\title{Quantum Effects in the Acoustic Plasmons of Atomically-Thin Heterostructures}

\author{A.~Rodr\'{\i}guez~Echarri}
\affiliation{ICFO-Institut de Ciencies Fotoniques, The Barcelona Institute of Science and Technology, 08860 Castelldefels (Barcelona), Spain}
\author{Joel~D.~Cox}
\affiliation{ICFO-Institut de Ciencies Fotoniques, The Barcelona Institute of Science and Technology, 08860 Castelldefels (Barcelona), Spain}
\author{F.~Javier~Garc\'{\i}a~de~Abajo}
\email{javier.garciadeabajo@nanophotonics.es}
\affiliation{ICFO-Institut de Ciencies Fotoniques, The Barcelona Institute of Science and Technology, 08860 Castelldefels (Barcelona), Spain}
\affiliation{ICREA-Instituci\'o Catalana de Recerca i Estudis Avan\c{c}ats, Passeig Llu\'{\i}s Companys 23, 08010 Barcelona, Spain}


\begin{abstract}
Recent advances in nanofabrication technology now enable unprecedented control over 2D heterostructures, in which single- or few-atom thick materials with synergetic opto-electronic properties can be combined to develop next-generation nanophotonic devices. Precise control of light can be achieved at the interface between 2D metal and dielectric layers, where surface plasmon polaritons strongly confine electromagnetic energy. Here we reveal quantum and finite-size effects in hybrid systems consisting of graphene and few-atomic-layer noble metals, based on a quantum description that captures the electronic band structure of these materials. These phenomena are found to play an important role in the metal screening of the plasmonic fields, determining the extent to which they propagate in the graphene layer. In particular, we find that a monoatomic metal layer is capable of pushing graphene plasmons toward the intraband transition region, rendering them acoustic, while the addition of more metal layers only produces minor changes in the dispersion but strongly affects the lifetime. We further find that a quantum approach is required to correctly account for the sizable Landau damping associated with single-particle excitations in the metal. We anticipate that these results will aid in the design of future platforms for extreme light-matter interaction on the nanoscale.
\end{abstract}

\maketitle

\section{Introduction}

The isolation of monolayer graphene \cite{NGM04} has stimulated extensive research efforts in two-dimensional (2D) materials, due in part to their unique electronic and optical properties, which are well-suited for use in compact, ultra-efficient photonic and opto-electronic devices \cite{XWX14}. Polaritons in 2D materials are particularly appealing in the field of nano-optics because they can confine external electromagnetic fields into extremely small volumes \cite{AND18}, enabling control of quantum and nonlinear optical phenomena on the nanoscale \cite{paper247}. Additionally, 2D polaritons are extremely sensitive to their surrounding environment, a property that renders them as good candidates for optical sensing \cite{KM06}, but also as enablers of new electro-optical functionalities when different atomically-thin materials are combined to form heterostructures \cite{GG13,paper283}.

Surface-plasmon polaritons (SPPs) are formed when light hybridizes with the collective oscillations of charge carriers at the interface between dielectric and conducting media \cite{E1969}, offering particularly strong confinement of electromagnetic energy down to truly nanometer length scales \cite{BDE03}. Noble metals are the traditional material platform used in plasmonics research, although they are difficult to actively tune and suffer from large ohmic losses \cite{MPG06,K15_2}. Graphene can help circumvent these limitations, as it supports highly-confined and long-lived plasmonic resonances that can be electrically modulated \cite{FAB11,paper196,FRA12}. In particular, the encapsulation of exfoliated monolayer graphene (MG) in hexagonal boron nitride (hBN) has been shown to dramatically improve the quality factor of plasmon resonances, with measured lifetimes of propagating modes $\sim0.5$\,ps at room temperature \cite{WLG15}, and even beyond 1\,ps at lower temperatures \cite{NMS18}. However, graphene plasmon studies have been so far limited to the terahertz and mid-infrared (mid-IR) spectral domains because the resonance energies of these excitations are severely constrained by the doping densities that can be sustained by the carbon layer, although some prospects have been formulated to extend their range of operation into the near-IR region \cite{paper235}.

Hybrid systems comprising noble metal layers and graphene potentially alleviate the limitations of plasmons in both of these materials by capitalizing on the appealing electrical tunability of graphene combined with the visible and near-IR plasmon resonances of noble metals. For example, tuning the damping of metal plasmons with electrostatically-gated graphene has been shown to enable a small degree of electrical tunability \cite{ECN12}, while this approach has been predicted to reach order-unity active modulation if a graphene layer is deposited on a thin metallic film \cite{paper277}. Still in the mid-IR, the proximity of a metal film in a heterostructure comprising hBN-encapsulated graphene and optically-thick metal surfaces has been recently demonstrated to push plasmon confinement down to a few atomic lengths outside the graphene sheet by rendering the plasmon dispersion acoustic, running close to the continuum of graphene intraband excitations \cite{LGA17,IND18}. Indeed, acoustic plasmons can result from the hybridization of modes in two neighboring metal surfaces \cite{DSA06}, or also in graphene and a metal surface, emerging as a low-energy branch in the energy-momentum dispersion diagram \cite{PAP11}. In contrast to the intuition that metal screening quenches the graphene response, an acoustic plasmon branch has been predicted to manifest itself even when graphene is directly deposited on the metal without a spacer \cite{PVP18}. For atomic-scale graphene-metal separations, the nonlocal nature of the optical response in both the metal and graphene layers plays an important role, demanding a rigorous theoretical treatment to accurately describe the dispersion and lifetime of acoustic plasmons \cite{DIG18}. The hydrodynamic model \cite{B1933,R1957,paper244,MRW14} has been extensively employed in this context, consisting in describing thin metal films within the framework of classical electrodynamics and introducing conduction electrons as a classical plasma \cite{RCW13,MCS13,DMC13,CPS13,DC17}. However, the hydrodynamical model neglects effects associated with the electronic band structure that can become important when optical confinement reaches the atomic length scale \cite{JL1975,HGD10,QXL15,RTW15,BS17}. First principles simulations have also been employed based on time-dependent density-functional theory to characterize the plasmons of metal films \cite{RG1984,PSC07,YJT11,LCA13,SAS15,SCP18,SCR18}, but they involve computationally expensive simulations that are difficult to extend to systems comprising heterostructures and lacking a single overall atomic periodicity. Nevertheless, a more rigorous study is still pending on the nonlocal effects that affect the plasmons supported by thin metal films and their hybridization with graphene modes.

Here, we explore the role played by quantum and finite-size effects in the acoustic plasmons resulting from hybridization of MG with few-atom-thick metallic films. We simulate the graphene and the metal within the random-phase approximation (RPA), incorporating in the metal phenomenological information on the main characteristics of the conduction electronic band structure, such as surface states, electron spill out, and a directional band gap associated with bulk atomic-layer corrugation. Comparison among different levels of approximation to the electron confinement potential, as well as to classical electromagnetic approaches, reveals that quantum and finite-size effects produce dramatic changes in the plasmon lifetime, which require a quantum description of the system to be properly accounted for. Remarkably, we find that a single metal monolayer can render the graphene plasmon dispersion acoustic, while the addition of more monolayers hardly modifies the dispersion but does change the lifetime. Also, the plasmon lifetime is strongly affected by Landau damping in the metal, which is automatically incorporated in the RPA description.  Our study and methodology are of general use for the description of plasmons in thin metal films and their interaction with nearby 2D materials.

\begin{figure*} 
\includegraphics[width=0.7\textwidth]{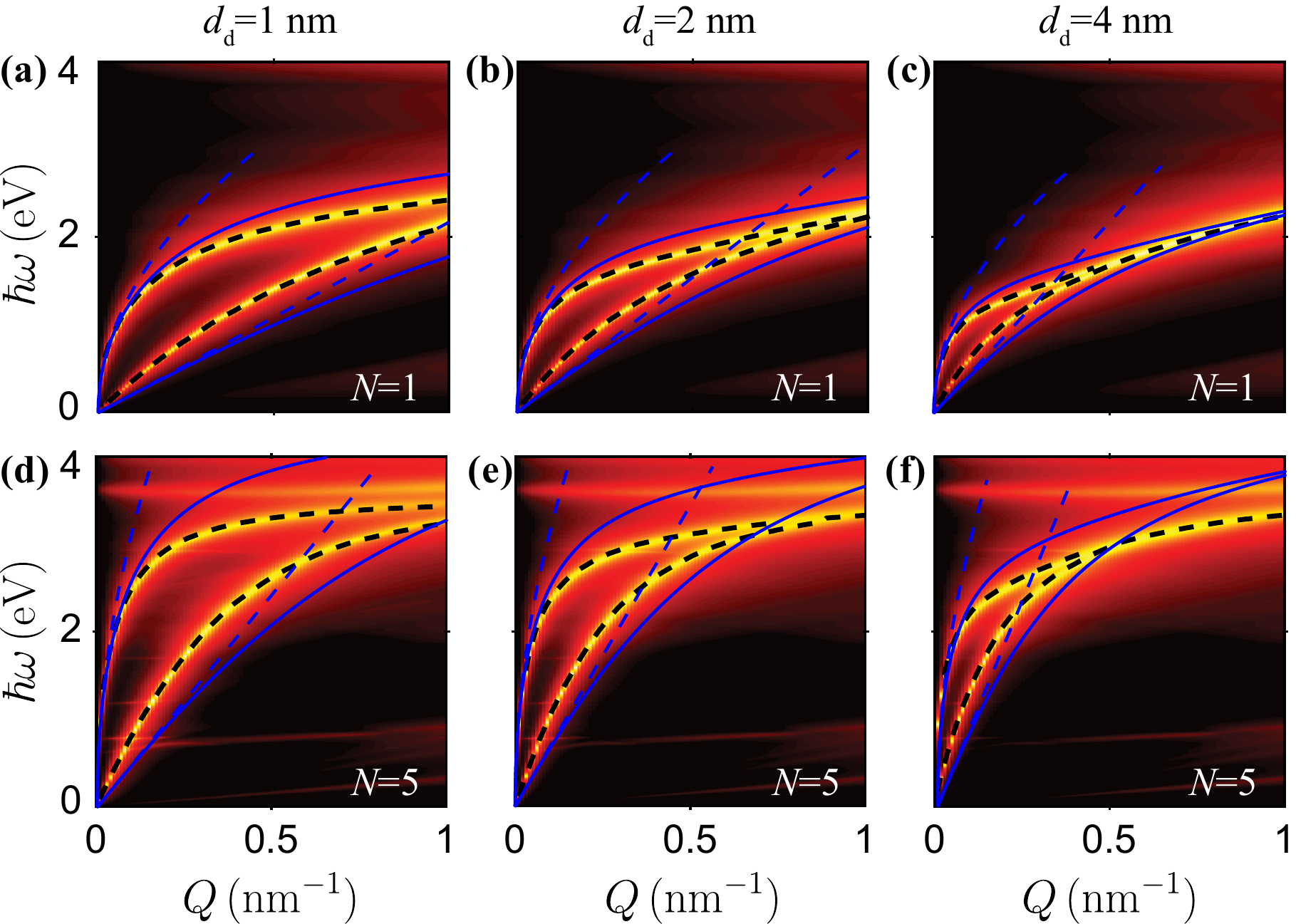}
\caption{\textbf{Plasmon interactions between two atomically thin silver films.} We consider symmetric structures consisting of two metallic films formed by either $N=1$ (a-c) or $N=5$ (d-f) Ag(111) atomic layers and separated by a dielectric (local permittivity $\epsilon=2.1$) of thickness $d_{\rm d}=1\,$nm (a,d), $d_{\rm d}=2\,$nm (b,e), or $d_{\rm d}=4\,$nm (c,f). QM results in the ALP model (color density plots) are compared with local dielectric theory based on the measured silver dielectric function \cite{JC1972} (black-dashed curves) and with the analytical expressions described in the main text (blue curves).}
\label{Fig1}
\end{figure*}

\section{Results and discussion}

Acoustic plasmons result from the repulsion between modes in closely spaced metallic films \cite{E1969} and their name derives from the linear relation between their frequency $\omega$ and in-plane wave vector $Q$, similar to acoustic waves. A tutorial description of this concept is provided by considering metal films of small thickness $d_{\rm m}$ and local permittivity $\epsilon_{\rm m}(\omega)$, the response of which can be approximated as that of a zero-thickness layer of 2D conductivity \cite{J99,paper254} $\sigma=[\ii\omega d_{\rm m}/(4\pi)]\,(1-\epsilon_{\rm m})$. When deposited on the planar surface of a dielectric of permittivity $\epsilon_{\rm d}$, the reflection coefficient for p polarization (i.e., the one associated with surface plasmons) reduces to $r=(1-\epsilon_{\rm d}+\xi)/(1-\epsilon_{\rm d}+\xi)$ (see Appendix), where $\xi=4\pi\ii\sigma Q/\omega$. Now, the plasmons of two films separated by a dielectric of thickness $d_{\rm d}$ and permittivity $\epsilon_{\rm d}$ are determined by the Fabry-Perot condition $r^2\ee^{-2Qd_{\rm d}}=1$, where the exponential represents the round trip for signal propagation across the dielectric spacer. Putting these elements together, we obtain two plasmon branches described to the expression $1-\epsilon_{\rm m}=(Qd_{\rm m})^{-1}\,F_\pm(Qd_{\rm d})$, where $F_\pm(x)=1+\epsilon_{\rm d}\,(1\mp\ee^{-x})/(1\pm\ee^{-x})$. In the Drude model for the metal conduction electrons, we can further approximate the metal permittivity as $\epsilon_{\rm m}\approx\epsilon_{\rm b}-\wp^2/\omega^2$, where $\wp$ is the bulk metal plasma frequency and $\epsilon_{\rm b}$ is a background permittivity accounting for polarization of inner electrons. This leads to a dispersion relation $\omega\approx\wp/\sqrt{\epsilon_{\rm b}-1+F_\pm/(Qd_{\rm m})}$, which we plot in Fig.\ \ref{Fig1} (blue-solid curves) for silver films (i.e., $\epsilon_{\rm b}=4$ and $\hbar\wp=9.17\,$eV, as obtained by fitting optical data \cite{JC1972}) of small thickness ($N=1$ and 5 Ag(111) atomic layers, with 0.236\,nm per layer) separated by a silica film ($\epsilon_{\rm d}=2.1$) of varying thickness $d_{\rm d}$. We find an upper plasmon branch and an acoustic plasmon at lower energies. In the small $Q$ limit, the above expressions can be further approximated as $\omega/\wp\approx\sqrt{Qd_{\rm m}}$ for the upper branch (i.e., with a characteristic $\omega\propto\sqrt{Q}$ dependence) and $\omega/\wp\approx Q\sqrt{d_{\rm d}d_{\rm m}/(2\epsilon_{\rm d})}$ for the acoustic plasmon branch (Fig.\ \ref{Fig1}, blue-dashed curves). This simple tutorial model is in reasonable agreement with a local dielectric description of the system (Fig.\ \ref{Fig1}, black-dashed curves) and it correctly captures the increasing degree of mode repulsion as $d_{\rm d}$ is reduced. In this work, we produce a more realistic quantum-mechanical simulation (see below), which we anticipate to yield a dispersion relation in remarkably excellent agreement with the above picture (Fig.\ \ref{Fig1}, color density plots, representing the loss function ${\rm Im}\{R\}$, where $R$ is the reflection coefficient of the metal-dielectric-metal structure for p polarization). The above discussion can be readily extended to metal-graphene and double-layer-graphene structures, yielding equally good agreement for the dispersion relations \cite{PAP11,PAP12}. Nevertheless, as we show below, an accurate calculation of the acoustic plasmon lifetimes is not possible without incorporating quantum-mechanical elements in the description of the system.

We are interested in studying the plasmons supported by hybrid planar films comprising an atomically thin metal layer and MG. Due to translational symmetry along the plane of the structures, each plasmon oscillating at an optical frequency $\omega$ can be characterized by a well-defined in-plane wave vector $\Qb$, so that its associated electromagnetic field depends on time and in-plane coordinates $\Rb=(x,y)$ just through an overall factor $\ee^{\ii\Qb\cdot\Rb-\ii\omega t}$, which is implicitly understood in what follows. We express the response of the hybrid film in terms of reflection and transmission coefficients for each of its constituting layers in a Fabry-Perot fashion (see Appendix). The calculation is simplified by the fact that the wavelengths of the plasmons under consideration are small compared with the light wavelength at the same frequency, therefore allowing us to work in the quasistatic limit, in which s-polarization components do not contribute. 

We describe graphene through its RPA conductivity \cite{WSS06,HD07} (see Appendix), which, in virtue of the van der Waals nature of its binding to the surrounding materials, should not be affected by the electronic properties of the latter, apart from some possible doping associated with carrier transfer. This level of description has been shown to excellently describe the optical response of graphene down to atomic length scales \cite{paper249}. In the main text we present results for graphene directly deposited on metal, while in the Appendix we provide additional calculations assuming a hBN intermediate layer, which is treated as a local, anisotropic dielectric film (see Appendix); this approach should capture the main modifications produced by the hBN layer on the response of the structure, which mainly consist of a featureless dielectric screening, accompanied by the signatures imprinted by its mid-IR phonon polaritons.

\begin{figure*}
\includegraphics[width=1\textwidth]{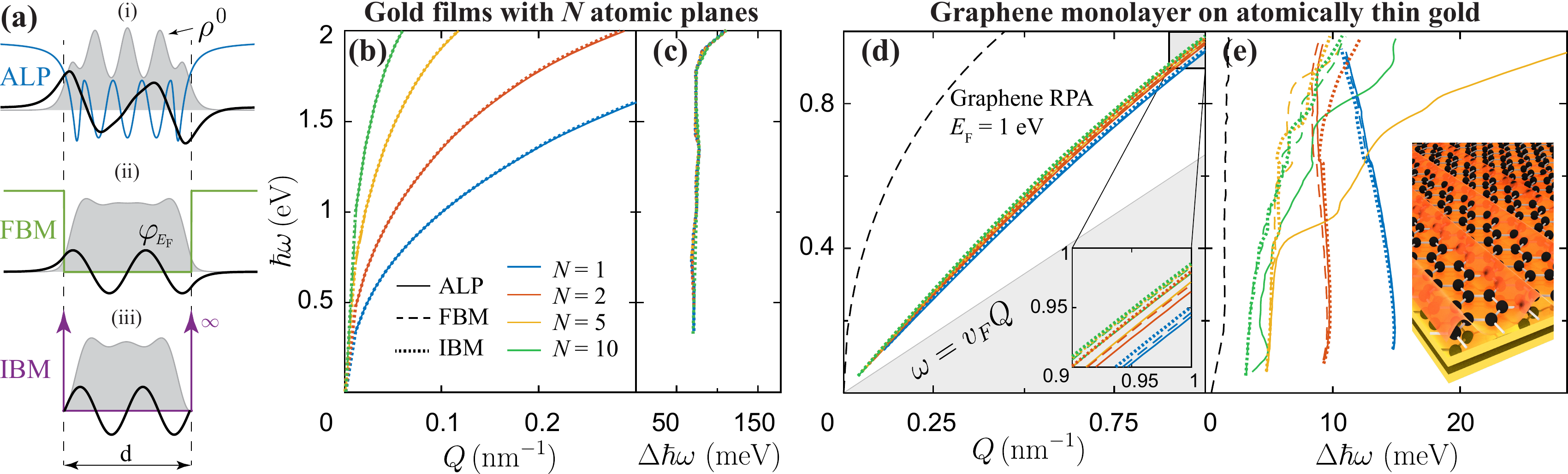}
\caption{\textbf{Quantum-mechanical models for plasmons in atomically-thin gold films and graphene-gold heterostructures.} {\bf (a)} Schematic illustration of several quantum-mechanical models used to describe a thin metallic film: atomic layer potential (ALP), finite barrier model (FBM), and infinite barrier model (IBM). The out-of-plane confinement potential is indicated by colored curves, while the unperturbed electronic charge density $\rho^{0}$ is shown by filled gray curves and a wave function of normal energy at the Fermi level by black curves. {\bf (b)} Plasmon dispersion relation of films comprising $N$ atomic Au(111) layers (0.236\,nm thickness per layer). Nearly indistinguishable results are obtained from the ALP, FBM, and IBM models. {\bf (c)} Full width at half maximum (FWHM) for the curves in (a). {\bf (d)} Low-energy acoustic plasmon dispersion relation of monolayer graphene (MG) deposited on the films considered in (b,c). The plasmon dispersion of isolated MG (dashed curve) and the intraband damping region $\omega \leq \vF Q$ (shaded area) are shown for reference. {\bf (e)} FWHM of the acoustic plasmons in (d). Curves for different models and gold-film thicknesses in (b)-(e) follow the legends of (b). Graphene is modeled in the RPA with Fermi energy $\EF=1\,$eV and an intrinsic lifetime of $500\,$fs in all cases.}
\label{Fig2}
\end{figure*}

The response of the metal layer is however sensitive to the proximity of the graphene. As we show below, this demands an accurate description of its nonlocal response. To this end, we adopt a QM model for the metal layer consisting in calculating its transmission and reflection coefficients within the RPA. Full details of the formalism are provided in the Appendix. In brief, the response of valence electrons is incorporated by calculating the non-interacting RPA susceptibility $\chi^0$ from the one-electron wave functions, while polarization of inner bands is accounted for through a local screened interaction. We further assume in-plane translational symmetry, so that the valence electron wave functions can be written as $\ee^{\ii\kparb\cdot\Rb}\varphi_j(z)$, labeled by the 2D in-plane electron wave vector $\kparb$ and the out-of-plane band index $j$. The wave function component $\varphi_j(z)$ is obtained upon solution of the one-dimensional (1D) Schr\"odinger equation along the out-of-plane direction, specified for a confining potential $V(z)$. We focus on gold and silver, for which a choice of effective mass $m^*=\me$ is appropriate, and more precisely, we apply this procedure to films consisting of a finite number $N$ of either Au(111) or Ag(111) atomic layers (metal thickness $\dm=Na_s$, with interlayer spacing $a_s\approx0.236\,$nm). The results of this procedure depend on the specific choice of potential $V(z)$, for which we use three different levels of approximation:
\begin{itemize}
\item {\it Atomic-layer potential (ALP).} We adopt from Ref.\ \cite{CSE99} a model potential that incorporates a harmonic corrugation in the bulk region and a smooth density profile at the surface, with parameters such that several important features of the electronic structure are correctly reproduced in the semi-infinite metal limit: the work function, the surface-projected bulk gap, and the position of the surface states relative to the Fermi level (see Fig.\ \ref{S1} in the Appendix, where the semi-infinite limit is approached for $N\sim100$). This potential therefore incorporates phenomenological information in a realistic fashion on (1) the out-of-plane quantization of electronic states, (2) the spill out of the electron wave functions beyond the metal film edges, and (3) the surface-projected electronic band gap produced by the bulk atomic-plane periodicity. Actually, this potential also describes surface and image states \cite{CSE99}, and therefore should realistically account for electron spill out effects. In practice, we designate the positions of the thin film surfaces as $z=0$ and $z=d_{\rm m}$, so that the first and last atomic planes are located at $z=a_s/2$ and $z=\dm-a_s/2$, respectively (see panel (i) in Fig.\ \ref{Fig2}a).
\item {\it Finite-barrier model (FBM).} Neglecting the atomic periodicity of the metal, we assign the potential $V(z)=V_0$ ($<0$) within the film ($0<z<d_{\rm m}$) and $V(z)=0$ otherwise (see panel (ii) in Fig.\ \ref{Fig2}a), extracting the potential barrier depth $V_0$ from the ALP for each metal. This choice of $V_0$ should provide the correct offset for the obtained electron energies. The FBM describes (1) 1D quantization of electronic states and (2) electron spill out beyond the metal film edges. 
\item {\it Infinite-barrier model (IBM).} To further simplify the model, we consider an infinite potential well such that $V(z)=V_0$ when $0<z<d_{\rm m}$ and $V(z) \to \infty$ elsewhere. The IBM accounts only for the quantization of electronic states along the film confinement direction.
\end{itemize}

In Fig.\ \ref{Fig2}b we show the SPP dispersion of self-standing gold films as predicted in the IBM, FBM, and ALP models, indicated by dotted, dashed, and solid curves, respectively, for gold films consisting of $N=1$-10 atomic layers. In each case we plot the frequency corresponding to the maximum of the loss function ${\rm Im}\{r_{\rm m}\}$, where $r_{\rm m}$ is the reflection coefficient for p polarization (see Appendix) at a given in-plane optical momentum $Q$. We obtain excellent agreement among the plasmon dispersions described by each choice of binding potential $V(z)$ within the QM model, even down to atomic-monolayer gold. We note that the dispersion curves predicted in the ALP are slightly redshifted with respect to those of the FBM, which in turn are redshifted relative to the dispersion of the IBM. Presumably, this redshift is related to electron spill out \cite{EBN12,paper244}, although an interplay between spill out and interband polarization is known to control the actual sign of the plasmon shift for small wave vectors \cite{L93}. The plasmon lifetime in the gold film is characterized by its spectral full width at half maximum (FWHM) $\Delta\omega$ plotted in Fig.\ \ref{Fig2}c, as determined from the linewidth of the ${\rm Im}\{r_{\rm m}\}$ spectral curves; the results, which are rather independent of the choice of potential model, lie near the intrinsic phenomenological width $\hbar\gamma=71$\,meV introduced in the RPA formalism for gold, a value taken from a Drude model fit of optical data \cite{JC1972}. We find a qualitatively similar behavior in silver films, but with smaller $\Delta\omega$ than in gold (see Fig.\ \ref{S2}a in the Appendix).

By depositing doped MG on the thin metal films, we introduce low-energy acoustic plasmon modes in the dispersion relation, for which metal screening confines light in the region between the graphene and the metal \cite{ANG17}. In Fig.\ \ref{Fig2}d we plot the calculated acoustic-plasmon dispersion relations for a graphene Fermi energy $\EF=1$\,eV obtained by using the three QM models considered in Fig.\ \ref{Fig2}b to account for the metal. Interestingly the acoustic plasmons show a dispersion that is rather independent of metal thickness down to $N=1$, apart from a minor shift towards the graphene intraband region ($\omega\leq\vF Q$, where $\vF\approx10^6\,$m/s is the Fermi velocity in graphene) with increasing $N$. A single atomic metal layer is thus capable of pushing the graphene plasmon dispersion (dashed curve) toward this region and render it acoustic. The associated FWHM (Fig.\ \ref{Fig2}e) of the acoustic plasmon is considerably reduced compared to its higher-energy bulk counterpart (Fig.\ \ref{Fig2}c) at similar wave vectors although (unlike the bulk branch the acoustic plasmon linewidth) it strongly depends on the metal thickness and the model used for the potential $V(z)$. It is remarkable that acoustic plasmons exist even by directly depositing the graphene on the metal without spacing, which confirms a recent prediction of this effect \cite{PVP18}, defying the intuition that metal screening can quench the graphene response. At low mid-IR frequencies, the acoustic plasmons are predicted to exhibit smaller lifetimes as the film thickness increases; thicker metal films are thus providing more efficient screening and less relative weight of the field inside the metal, where the intrinsic inelastic rate is larger than in the graphene alone (cf. dashed curve, corresponding to the assumed intrinsic lifetime $\tau=500\,$fs in graphene). At higher energies, the results become more involved, as metal screening is less effective, and the FWHM depends more strongly on the model used for the potential. We remark that the ALP model, which is expected to provide the most accurate description because it incorporates several phenomenological features of the electronic band structure, leads to generally higher damping than the other models.

\begin{figure*}
\includegraphics[width=1\textwidth]{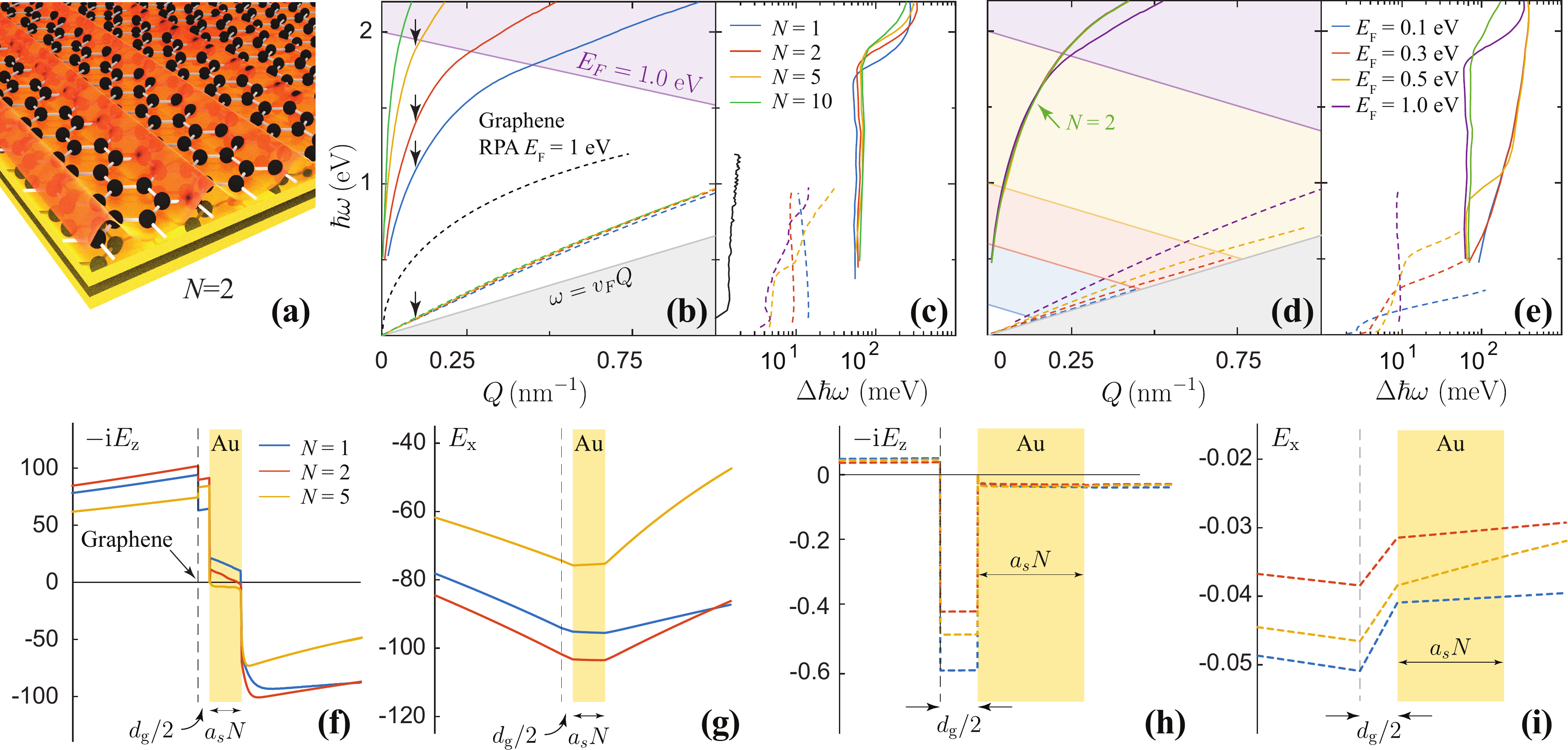}
\caption{\textbf{Thickness and doping dependence of plasmons in graphene-gold films.} {\bf (a)} Schematic illustration of a MG-gold film heterostructure containing $N=2$ atomic Au(111) layers. {\bf (b,c)} Dispersion relation (b) and FWHM (c) of high-energy (solid curves) and acoustic (dashed curves) plasmons for a graphene Fermi energy $\EF=1$\,eV and different gold-film thicknesses (see legend for $N$ in (c)). {\bf (d,e)} Plasmon dispersion (d) and FWHM (e) for $N=2$ and various Fermi energies (see legend for $\EF$ in (e)). {\bf (f-i)} Spatial distribution of the in-plane ($E_x$) and out-of-plane ($-\ii E_z$) components of the plasmon electric field for various metal thicknesses (see legend in (f)) at the wave vector indicated by the black arrows in (b) (i.e., $Q=0.1\,$nm$^{-1}$). Solid and dashed curves represent the fields for the high-energy and acoustic plasmons, respectively (see (b)). The graphene layer is approximated as a zero-thickness film (vertical dashed lines, separated $d_{\rm g}/2$ from the metal, where $d_{\rm g}=0.33\,$nm is the nominal graphene thickness); the near fields of the acoustic plasmons (h,i) are mainly concentrated in the graphene-metal region, whereas the higher-energy plasmons (f,g) exhibit more delocalized field profiles. We describe graphene in the RPA in all plots.}
\label{Fig3}
\end{figure*}

In Fig.\ \ref{Fig3} we study the full plasmon dispersion of the MG-gold film heterostructure, which is schematically illustrated in Fig.\ \ref{Fig3}a for a system comprised of only two atomic gold layers (see also Fig.\ \ref{S3} in the Appendix for contour plots of the loss function ${\rm Im}\{R\}$). Fig.\ \ref{Fig3}b indicates that, while the dispersion of the acoustic plasmon is only marginally influenced by the film thickness, the higher-energy plasmon retains a similar dependence as that of the isolated film considered in Fig.\ \ref{Fig2}b. At high energies approaching the interband damping regime in graphene ($\hbar \omega \geq2\EF-\hbar\vF Q$), the high-energy plasmon dispersion undergoes a slight redshift, accompanied by an increase in the effective plasmon damping (see Fig.\ \ref{Fig3}c).

It is remarkable that the presence of a single atomic metal layer ($N=1$) is sufficient to support plasmons (see Figs.\ \ref{Fig2}d and \ref{Fig3}b), with a group velocity at low frequencies $v_{\rm g}\approx1.7\times10^6\,$m/s for $\EF=1\,$eV that is nearly unchanged by the addition of further atomic metal layers. These modes can be modulated by varying the Fermi energy $\EF$ in MG, with relatively small damping in the $\hbar\omega\lesssim\EF$ region, while strong attenuation is produced by interband transitions at higher energies (Fig.\ \ref{Fig3}d,e).

\begin{figure*}
\includegraphics[width=1\textwidth]{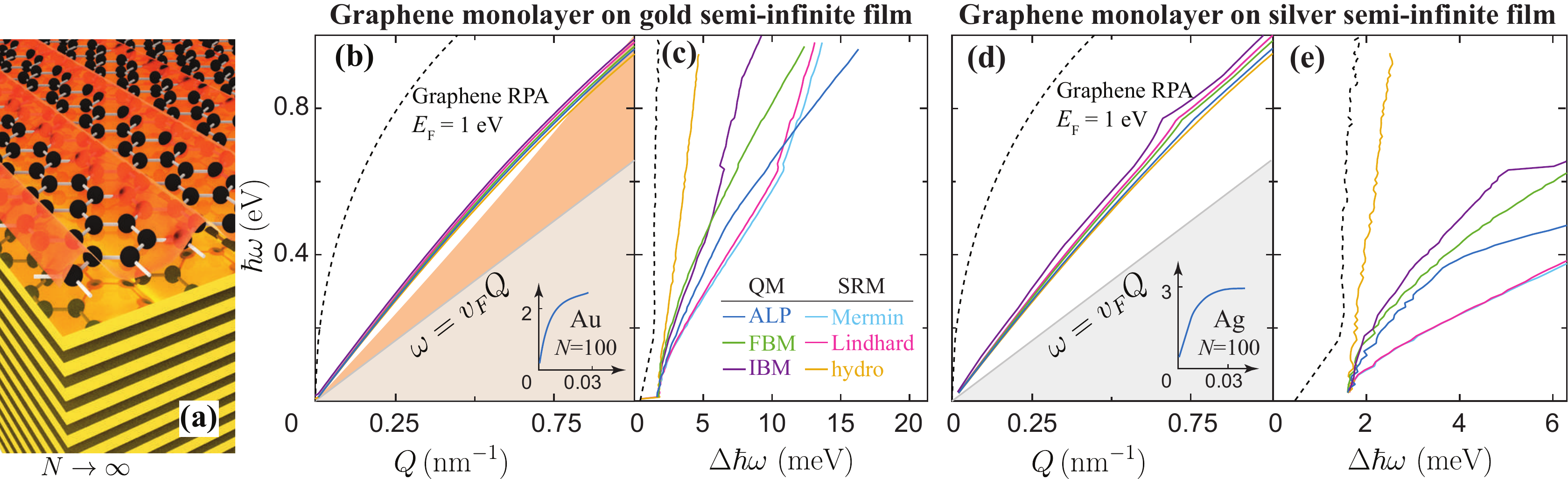}
\caption{\textbf{QM vs classical description of acoustic plasmons in graphene-metal hybrid systems.} {\bf (a)} We consider MG on optically thick gold or silver films. {\bf (b,c)} Dispersion relation (b) and associated FWHM (c) for MG-gold with $\EF=1$\,eV doping, as obtained by using different QM and classical (specular-reflection model, SRM) approaches to describe the response of gold (see legend in (c) and details in the main text). {\bf (d,e)} Same as (b,c) when gold is replaced by silver. We model graphene in the RPA in all cases. The film thickness is infinite in the classical SRM, whereas we take $N=100$ (i.e., 23.6\,nm) films in the QM approach. The orange shaded area in (b) denotes the bulk electron-hole pair region in gold (see main text). The lower-right insets to (b,d) show the dispersion of the high-energy surface plasmon.}
\label{Fig4}
\end{figure*}

Examination of the effective damping rates in Figs.\ \ref{Fig2}e and \ref{Fig3}c indicates that an increase in film thickness can reduce the plasmon linewidth, owing to more effective screening by the metal layer. We are thus prompted to consider acoustic plasmons in the MG-gold film heterostructure in the limit of a semi-infinite metal layer, which is schematically illustrated in Fig.\ \ref{Fig4}a. Our QM treatment of a thin film can also describe the response of a semi-infinite film if we approach the $N\rightarrow\infty$ limit, which in practice is reached for $N\sim100$ atomic layers (see Fig.\ \ref{S1} in the Appendix).

We find it instructive to compare the above QM descriptions of the metal to classical approaches that are extensively used in recent literature. In particular, we consider the so-called specular-reflection model \cite{RM1966} (SRM), also known as semiclassical infinite barrier model \cite{FW1984}, which allows us to express the response of a semi-infinite material in terms of the nonlocal dielectric function of the bulk, while a straightforward generalization can deal with arbitrary shapes \cite{paper119}. Following the detailed prescription presented in the Appendix, we apply this model to three different choices for the bulk dielectric response function: the hydrodynamical model \cite{B1933,R1957}, the Lindhard dielectric function \cite{L1954,HL1970}, and the Mermin prescription \cite{M1970}. The bulk hydrodynamic model combined with the SRM coincides with the hydrodynamic model for finite geometries, which has been extensively used to discuss nonlocal effects in nanostructured metals \cite{paper119,MGS10_2,MRW14}; it incorporates nonlocal effects though the hydrodynamic pressure, as well as a phenomenological damping rate $\gamma$. The Lindhard dielectric function is used here with a correction intended to account for d-band screening \cite{paper119} (see Appendix), as well as a damping rate $\gamma$ effectively introduced by replacing $\omega$ by $\omega+\ii\gamma/2$ in the Lindhard formula. The Mermin prescription corrects the Lindhard formula in order to preserve local conservation of electron density during damping processes \cite{M1970}.

In Fig.\ \ref{Fig4}b,c we compare the performance of the classical (i.e., SRM using hydrodynamic, Lindhard, and Mermin dielectric functions) and QM (i.e., RPA with ALP, FBM, and IBM potentials) approaches for MG deposited on an optically-thick gold surface. The dispersion relation is very similar in both cases. In contrast, we find significant discrepancies in the linewidth: the classical approach overestimates plasmon broadening when using the Lindhard and Mermin dielectric functions, but it produces a severe underestimate in the hydrodynamical model, presumably because the latter does not account for the efficient mechanism of Landau damping associated with decay into metal electron-hole pair (e-h) excitations; indeed, the bulk e-h region in gold (orange area in Fig.\ \ref{Fig4}b, defined by $\omega\le (\hbar/m^*)(Q^2/2+Qv_{\rm F}^{\rm Au})$, where the gold Fermi velocity $v_{\rm F}^{\rm Au}\approx1.4\times10^6\,$m/s exceeds by 40\% that of graphene) overlaps the plasmon dispersion for $\hbar\omega\gtrsim0.8\,$eV, where $\Delta\omega$ takes relatively large values. In addition to this, the finite damping introduced in the model and the lack of translational symmetry along the surface normal direction effectively extend coupling to the e-h region toward low in-plane wave vectors, thus increasing its overlap with the plasmon. Remarkably, the plasmon damping predicted by the QM approach using the IBM potential agrees well with the SRM approach using the Lindard and Mermin dielectric functions; the main difference between them lies in the neglect of quantum interference between outgoing and surface-reflected electron components in the SRM (see Appendix), which seems to play a minor role. Finally, in Fig.\ \ref{Fig4}c,d we consider optically-thick silver surfaces, which exhibit qualitatively similar behavior as gold, although the reduced intrinsic inelastic damping of the argent metal ($\hbar\gamma=21$ meV) results in smaller plasmon linewidths.

We obtain qualitatively similar results when MG is separated from the metal by a thin layer of hBN (see Figs.\ \ref{S4} and \ref{S5} in the Appendix). The latter introduces sharp spectral features characterized by avoided crossing between the plasmons and the mid-IR photons of the insulator. Additionally, the MG-metal interaction is reduced relative to the structures without hBN because of their larger separation, therefore yielding low-energy plasmon bands with a less acoustic character. This effect is clearly observed when comparing contour plots of the loss function ${\rm Im}\{R\}$ in both types of structures (cf. Figs.\ \ref{S3} and \ref{S6} in the Appendix). Interestingly, although the acoustic plasmon dispersion depends strongly on the spacer thickness and level of graphene doping, it is not so sensitive to the permittivity of the dielectric spacer (Fig.\ \ref{S7} in the Appendix).

\begin{figure*}
\includegraphics[width=0.7\textwidth]{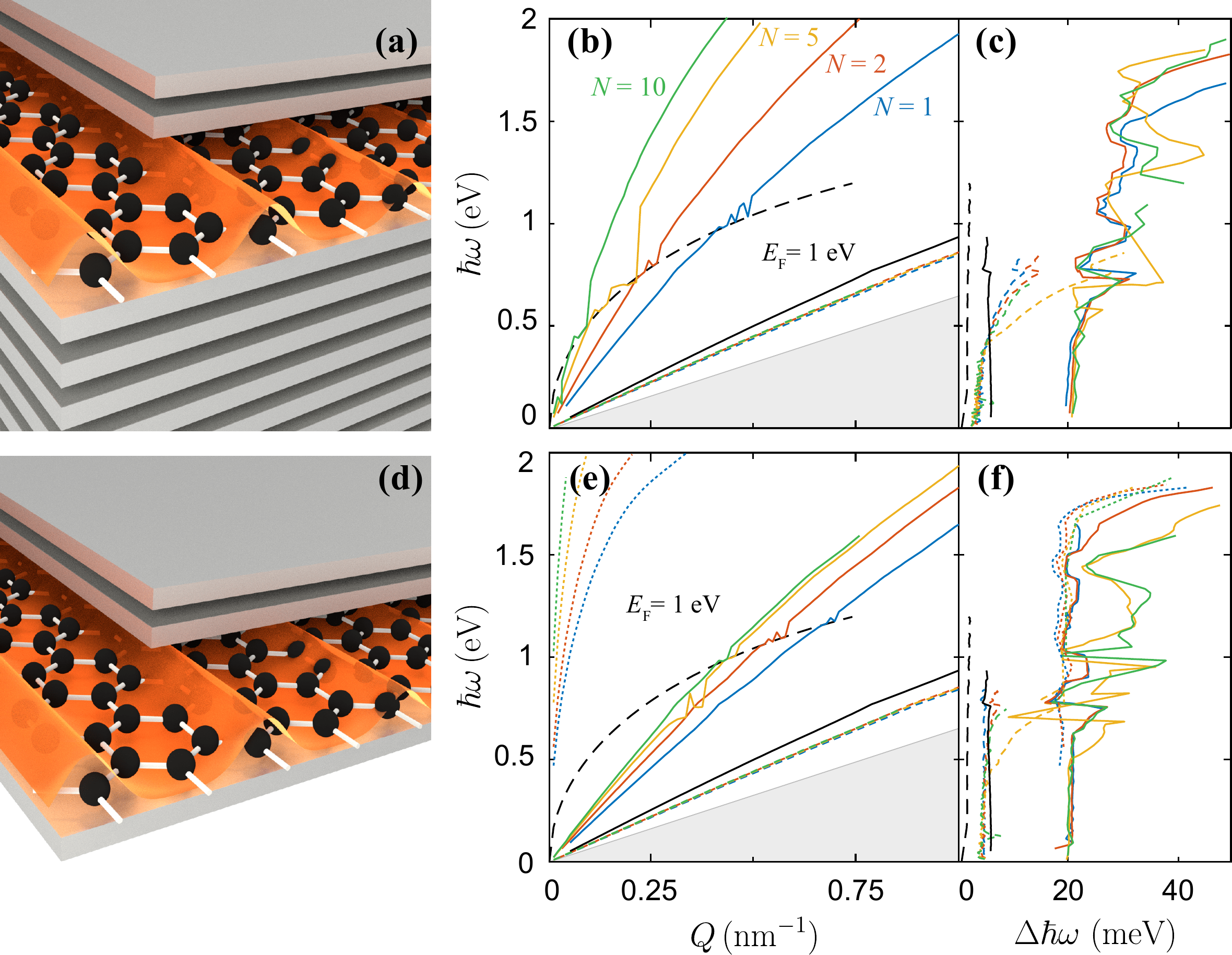}
\caption{\textbf{Plasmons in metal-graphene-metal structures.} We consider graphene directly embedded in silver without dielectric spacers. {\bf (a)} Illustration of a structure consisting of a finite-thickness silver film ($N=2$ Ag(111) atomic layers in the sketch) deposited on a MG sheet, which in turn rests on an optically thick Ag(111) surface. {\bf (b,c)} Plasmon dispersion (b) and FWHM (c) of the structure in (a) for various upper-film thicknesses (see color-coded labels $N$). The graphene doping level is $\EF=1\,$eV. Three different plasmon branches are observed in each structure: two acoustic ones (lower dashed curves and upper solid curves) and a high-energy plasmon (not shown) that is nearly overlapping the vertical energy axis. {\bf (d-f)} Same as (a-c), with the bottom metal surface replaced by a monolayer Ag(111) sheet; here we show the upper-energy plasmon (dotted curves), which is no longer separated from the vertical axis. As a reference, we show the plasmon dispersion for isolated MG (black long-dashed curves) and the acoustic plasmon observed in the structures without the upper film (black solid curves).}
\label{Fig5}
\end{figure*}

An interesting scenario arises when graphene is fully embedded in metal. We can easily simulate this type of structure through a trivial extension of the Fabry-Perot approach discussed in the Appendix. The results (Fig.\ \ref{Fig5}, for graphene doped to $EF=1\,$eV and embedded in silver) reveal an interplay between three different plasmon modes provided by the graphene and each of the two metal regions. In particular, we find again an acoustic plasmon that is mainly localized in the graphene, now pushed even further toward the intraband transition region. Additionally, a higher-energy acoustic plamon emerges, and this mode becomes more acoustic and confined when one of the lower metal region is a single atomic layer (cf. Figs.\ \ref{Fig5}b and \ref{Fig5}c) because the plasmons in the two metal films are then closer in energy, so they undergo stronger interaction, similar to what we observed in Fig.\ \ref{Fig1}. This second acoustic plasmon exhibits larger tunability with the number of metal atomic layers, although it is more lossy than the one dominated by graphene, as it has more weight in the metal, which is taken to have a lower intrinsic inelastic lifetime (31\,fs for silver \cite{JC1972}) than graphene (500\,fs). Similar conclusions are observed for lower graphene doping (see Fig.\ \ref{S8} in the Appendix, where we consider $\EF=0.2$\,eV).

\section{Conclusion}

The acoustic plasmons supported by monolayer graphene when it is deposited on noble metal surfaces are strongly influenced by the metal optical response. We have identified quantum finite-size effects in the optical response of ultrathin films and semi-infinite metal layers that impact the acoustic plasmons formed by interaction with doped graphene. Remarkably, a single atomic metal layer is capable  of rendering the graphene plasmons acoustic, displaying a group velocity $\sim1.6\,\vF$ for a doping $\EF=1\,$eV that is nearly insensitive to the addition of more metal layers. The level of metal screening, which influences the plasmon lifetimes, is however strongly dependent on metal thickness. We further reveal the important contribution of electron-hole pairs in the metal to the plasmon damping, a fair description of which can only be achieved by incorporating details of the metal electronic band structure, electron spill out, and surface-projected bulk gaps, such as we do in this work using a computationally efficient quantum approach. The present study could be improved by using a first principles estimate of the internuclear distance in the MG-metal interface, as well as by introducing the effect of in-plane atomic corrugation, although these are computationally demanding calculations that we expect to produce only minor corrections. We anticipate that the results presented here can elucidate the role of quantum and finite-size effects in acoustic plasmons, enabling a fundamental understanding of the damping and extreme spatial confinement associated with these excitations.

\acknowledgments

This work has been supported in part by the Spanish MINECO (MAT2017-88492-R and SEV2015-0522), the ERC (Advanced Grant 789104-eNANO),  the European Commission (Graphene Flagship 696656), AGAUR (2017 SGR 1651), the Catalan CERCA Program, and Fundaci\'o Privada Cellex. A.R.E. acknowledges a grant cofounded by the Generalitat de Catalunya and the European Social Fund FEDER.

\appendix

\section{Optical response of a thin film heterostructure}
We consider a multilayer hybrid film consisting of MG deposited on a metal film, from which it is separated by a thin dielectric layer. The optical response is expressed in terms of the reflection ($r_i$) and transmission ($t_i$) coefficients of the graphene layer ($i=$\,g), the dielectric spacer ($i=$\,d), and the metal film ($i=$\,m) by describing the heterostructure as a Fabry-Perot resonator characterized by the total reflection and transmission coefficients \cite{LFL17}
\begin{align}
R&=r_{\rm gd}+ \frac{t_{\rm gd} t_{\rm dg} r_{\rm m}}{1-r_{\rm dg}r_{\rm m}}, \nonumber\\
T&=\frac{t_{\rm gd} t_{\rm m}}{1-r_{\rm dg}r_{\rm m}}, \nonumber
\end{align}
where the doubly-indexed coefficients combining graphene and the dielectric layer are defined as
\begin{align}
r_{ij}&=r_i+\frac{t_i^2 r_j}{1 - r_i r_j}, \label{rij}\\
t_{ij}&=\frac{t_i t_j}{1 - r_i r_j}. \nonumber
\end{align}
This prescription is convenient because we can use coefficients calculated for a surrounding vacuum by just taking the graphene-dielectric and dielectric-metal spacings as consisting of a zero-thickness vacuum layer. In the quasistatic limit here considered (see below), only p-polarization coefficients make a nonzero contribution. Explicit values of these coefficients are given below for graphene and hBN, accompanied by a quantum-mechanical approach to deal with the metal film, which we compare with a nonlocal classical solution in the thick-metal limit. While in the main text we disregard the dielectric spacer by imposing $r_{\rm d}=0$ and $t_{\rm d}=1$, we provide in the Appendix complementary simulations including the effect of a thin hBN spacing layer between the graphene and the metal.

\section{Quantum-mechanical description of the metallic film}
We study the linear optical response of a self-sustained metal film of thickness $\dm$ contained in the $0<z<\dm$ region within the RPA \cite{PN1966}, for which we construct the non-interacting susceptibility $\chi^0$ in terms of one-electron wave functions, which are in turn calculated as the solutions of an effective potential $V(z)$. For simplicity, we assume translational invariance along the in-plane directions (coordinates $\Rb=(x,y)$) and adopt different models for the out-of-plane potential profile (see Fig.\ \ref{Fig2}a). This allows us to express the optical response in terms of components of well-defined in-plane parallel wave vector $\Qb$ and frequency $\omega$, so we assume an implicit dependence on these variables, as well as a multiplicative factor $\ee^{\ii(\Qb\cdot\Rb-\omega t)}$. For example, the potential in real space-time has the form $\phi(\rb,t)=(2\pi)^{-3}\int d^2\Qb\int d\omega \,\ee^{\ii(\Qb\cdot\Rb-\omega t)}\,\phi(\Qb,z,\omega)$. Additionally, we ignore retardation effects because the wavelengths of the plasmons under investigation are much smaller than the light wavelength, and consequently, we study the response using electrostatic potentials. Specifically, we consider an external ({\it incident}) potential $\phi^{\rm ext}(z)=(2\pi/Q)\ee^{-Q|z-z_0|}$ representing a source located below the film at $z=z_0$ (in practice we take $z_0=0$). The reflection coefficient is defined as the ratio of induced to external potentials $r_{\rm m}=1-\phi(z_0)/\phi^{\rm ext}(z_0)$, where an overall minus sign is introduced to make this definition coincide with the quasistatic limit of the Fresnel coefficient for p polarization. Likewise, the transmission coefficient is defined as $t_{\rm m}=\phi(z_1)/\phi^{\rm ext}(z_0)$, where the {\it transmitted} potential is evaluated at a position $z_1$ right above the film (we take $z_1=\dm$).

The response of inner shells is accounted for by introducing a homogeneous film of local background permittivity $\epsb(\omega)$ contained in the $0<z<\dm$ region, so that it extends half an atomic-layer spacing beyond each of the two outermost atomic planes (see ALP below). We note that $\epsb$ can take relatively large values in noble metals within the visible and near-IR spectral ranges (e.g., $\sim9$ in Au) due to d-band polarization. In practice, we obtain $\eps_{\rm b}(\omega)$ by subtracting from the experimentally measured metal dielectric function \cite{JC1972} $\eps_{\rm exp}(\omega)$ a Drude term, such that $\epsilon_{\rm b}(\omega)=\epsilon_{\rm exp}(\omega)+\omega_{\rm p}^2/\omega(\omega+\ii\gamma)$, where $\hbar\omega_{\rm p}=9.06\,$eV, $\hbar\gamma=0.071\,$eV for gold and $\hbar\omega_{\rm p}=9.17\,$eV, $\hbar\gamma=0.021\,$eV for silver (see Fig.\ \ref{S9} in the Appendix for plots of $\eps_{\rm exp}(\omega)$ and $\eps_{\rm b}(\omega)$).

We express the total potential as
\begin{align}
\phi(z)=\phi^{\rm ext}_{\rm b}(z)+\int dz'\, v_{\rm b}(z,z')\,\rho^{\rm ind}(z'),
\nonumber
\end{align}
where the $\phi^{\rm ext}_{\rm b}$ term accounts for the potential created by $\phi^{\rm ext}$ in the presence of the background slab of permittivity $\epsb$ (i.e., including the background response, but not the response of the conduction electrons), while the integral gives the contribution due to the induced-charge density $\rho^{\rm ind}(z')$ associated with disturbances in the conduction electrons. The latter is mediated by the screened interaction $v_{\rm b}$ (i.e., the Coulomb potential created at $z$ by a point charge at $z'$ oscillating with frequency $\omega$ in $\Qb$ space, including the effect of background polarization). In the absence of background polarization (i.e., for $\epsb=1$) we have $v_{\rm b}(z,z')=(2\pi/Q)\ee^{-Q|z-z'|}$, while in the presence of an $\epsb\neq1$ film we still find a closed-form expression by direct solution of Poisson's equation \cite{paper295}:
\begin{align}
v_{\rm b}(z,z')=v_{\rm b}^{\rm dir}(z,z')+v_{\rm b}^{\rm ref}(z,z'),
\nonumber
\end{align}
where 
\begin{widetext}
\begin{align}
v_{\rm b}^{\rm dir}(z,z')=\frac{2\pi}{Q}\,\ee^{-Q|z-z'|}\times
\left\{\begin{array}{ll}
1,  &z,z'\le0 \text{ or } z,z'>\dm \\
1/\epsb,  &0<z,z'\le\dm \\
0, & \text{otherwise}
\end{array} \right.
\nonumber
\end{align}
is the direct Coulomb interaction in each homogeneous region of space,
\begin{align}
v_{\rm b}^{\rm ref}(z,z')=g
\times
\left\{\begin{array}{ll}
(1-\epsb^2)\left(\ee^{2Q\dm}-1\right)\,\ee^{-Q(z+z')},  &\dm<z, z' \\ 
2\left[(\epsb+1)\ee^{-Q(z-z')}+(\epsb-1)\ee^{-Q(z+z')}\right],  & 0<z'\le\dm<z \\ 
4\epsb\;\ee^{-Q(z-z')},  & z'\le0\;\;\text{and}\;\;\dm<z \\ 
2\left[(\epsb+1)\ee^{Q(z-z')}+(\epsb-1)\ee^{-Q(z+z')}\right],  & 0<z\le\dm<z' \\ 
(1/\epsb)\big\{
(\epsb^2-1)\left[\ee^{-Q(z+z')}+\ee^{-Q(2\dm-z-z')}\right] & \nonumber\\
\quad\;\;\;
+(\epsb-1)^2\left[\ee^{-Q(2\dm+z-z')}+\ee^{-Q(2\dm-z+z')}\right]
\big\},  & 0<z,z'\le\dm \\ 
2\left[(\epsb+1)\ee^{-Q(z-z')}+(\epsb-1)\ee^{-Q(2\dm-z-z')}\right],  & z'\le0<z\le\dm \\ 
4\epsb\;\ee^{Q(z-z')},  &z\le0\;\;\text{and}\;\;\dm<z'\\ 
2\left[(\epsb+1)\ee^{Q(z-z')}
+(\epsb-1)\ee^{-Q(2\dm-z-z')}\right],  &z\le0<z'\le\dm\\ 
(1-\epsb^2)\left(1-\ee^{-2Q\dm}\right)\,\ee^{Q(z+z')},  &z, z'\le0
\end{array} \right.
\nonumber
\end{align}
\end{widetext}
accounts for the effect of reflections at the film surfaces, and we define
\begin{align}
g=\frac{(2\pi/Q)}{(\epsb+1)^2-(\epsb-1)^2\ee^{-2Q\dm}}. \nonumber
\end{align}
The $v_{\rm b}^{\rm dir}$ term captures the charge singularity in the interaction potential within each homogeneous region of space, whereas the addition of the $v_{\rm b}^{\rm ref}$ term guarantees the continuity of the potential and the normal displacement at the interfaces. Using this expression, and noticing that the external potential originates in a source at $z_0\le0$, we can readily write the external potential including the interaction with the background film as $\phi^{\rm ext}_{\rm b}(z)=v_{\rm b}(z,z_0)$.



Assuming linear response, we can express $\rho^{\rm ind}$ in terms of the susceptibility $\chi$ according to
\begin{align}
\rho^{\rm ind}(z)=\int dz'\chi(z,z')\,\phi^{\rm ext}_{\rm b}(z'),
\nonumber
\end{align}
where the external potential $\phi^{\rm ext}_{\rm b}$ driving the free electrons has been corrected by the direct background contribution as explained above. Using matrix notation with $z$ acting as an index and a dot indicating integration over this coordinate, we can write
\begin{align}
\chi=\chi^0\cdot\left(1-v_{\rm b}\cdot\chi^0\right)^{-1},
\label{chich0}
\end{align}
where $\chi^0$ is the non-interacting RPA susceptibility \cite{PN1966}. We use the well-known result \cite{HL1970}
\begin{align}
\chi^0(\rb,\rb') = \frac{2e^2}{\hbar} \sum_{ii'} \left(f_{i'}-f_{i}\right) \frac{\psi_i(\rb) \psi_i^*(\rb') \psi_{i'}^*(\rb) \psi_{i'}(\rb')}{\omega+\ii\gamma-(\en_i-\en_{i'}) }
\label{chi_0_eq}
\end{align}
for the full spatial dependence of this quantity in terms of the one-electron metal wave functions $\psi_i(\rb)$, where the factor of 2 accounts for spin degeneracy, $f_i$ is the occupation of state $i$ with energy $\hbar\en_i$, and $\gamma$ denotes a phenomenological inelastic damping rate \cite{HL1970,PN1966}. In order to exploit translational invariance in the film, we multiplex the state index as $i\rightarrow\{j,\kparb\}$, where $j$ labels eigenstates of the $z$-dependent out-of-plane 1D Schr\"odinger equation defined by the noted binding potential $V(z)$ with an effective mass $m^*$ (see below), while $\kparb$ runs over 2D in-plane electron wave vectors. The electron wave functions have therefore the form $\psi_i(\rb)=A^{-1/2}\ee^{\ii \kparb \cdot \Rb}\,\varphi_j(z)$, where $A$ is the film normalization area and $\varphi_j(z)$ are eigenstates of the 1D problem. Using these wave functions, we can recast Eq.\ (\ref{chi_0_eq}) as $\chi^0(\rb,\rb')=(2\pi)^{-2}\int d^2\Qb\,\ee^{\ii\Qb\cdot(\Rb-\Rb')}\,\chi^0(z,z')$, where
\begin{align}
\chi^0(z,z') = &\frac{e^2}{2\pi^2\hbar} \sum_{jj'} \int d^2\kparb\, \left(f_{j',|\kparb-\Qb/2|}-f_{j,|\kparb+\Qb/2|}\right)
\nonumber\\
&\times \frac{\varphi_j(z) \varphi_j^*(z') \varphi_{j'}^*(z) \varphi_{j'}(z')}{\omega+\ii\gamma-\left[\en_j-\en_{j'}+(\hbar/m^*)\kparb\cdot\Qb\right]}
\nonumber
\end{align}
is the quantity actually used in Eq.\ (\ref{chich0}). Here, we define occupation numbers $f_{j,\kpar}=\Theta(\EF-\hbar\en_j-\hbar^2\kpar^2/2m^*)$ that follow the Fermi-Dirac distribution at zero temperature for a Fermi energy $\EF$, where the rightmost term inside the step function describes a parabolic dispersion along in-plane directions with effective mass $m^*$ as well. We evaluate these expressions by expanding all quantities in sine-Fourier transform within an embedding infinite-potential box spanning a $\sim1\,$nm vacuum region on each side of the film. The quantities $\chi^0$, $v_{\rm b}$, and $\chi$ then become square matrices in this representation, so we operate with them using linear algebra techniques. We have further corroborated the accuracy of our numerical results by comparing with a real-space discretization in the $z$ coordinate, which further reduces spurious Gibbs' oscillations in the calculation of the near fields, although the sine basis generally converges with a smaller number of elements.

The average volumetric electron density in the film is $n=[2/(A\dm)]\, \sum_j\sum_{\kparb} f_{j,\kpar}$, where the factor of 2 is again due to spin degeneracy. Using the customary transformation $\sum_{\kparb} \to (2\pi)^{-2}A\int d^2\kparb$, we obtain the self-consistent expression for the Fermi energy
\begin{align}
\EF=\frac{\hbar}{M}\left[\frac{\pi \hbar d_{\rm m}}{m^*} n+\sum_{j=1}^M \en_j\right],
\nonumber
\end{align}
where $M$ denotes the highest band index for which $\hbar\en_M<\EF$. To correctly define $\EF$ for arbitrary thickness, we fit the charge density $n$ in the bulk limit ($d_{\rm m}\to\infty$) by imposing experimentally-measured values of $\EF$ relative to vacuum in gold ($-5.5$\,eV) and silver ($-4.65$\,eV). The choice of potential parameters \cite{CSE99} in the ALP further ensures that the binding energies of surface states agree with available experimental measurements \cite{PMM95} (see thickness-dependent band structure in Fig.\ \ref{S1} in the Appendix). Under these conditions we obtain effective charge densities $n=70.5\,$nm$^{-3}$ for gold and $n=59.6\,$nm$^{-3}$ for silver. For the sake of consistency, we use these values of $n$ in the calculations of $\chi^0$ presented in this work for all models of $V(z)$ (see Fig.\ \ref{Fig2}a), and we also assume the same density per layer for finite-thickness films.

\section{Nonlocal classical electromagnetic description of a semi-infinite metal film}
For the sake of comparison in the thick-film limit, we adopt the SRM \cite{RM1966,paper119,PSC07} to compute a nonlocal reflection coefficient $r_{\rm m}$ for a semi-infinite metal in the framework of classical electrodynamics. This model allows us to relate the surface response to the momentum- and frequency-dependent dielectric function of the metal $\epsilon_{\rm m}(q,\omega)$ under the assumption that conduction electrons undergo specular reflection at the surface without quantum interference between outgoing and reflected components. In practical terms, we calculate the reflection coefficient in the SRM as \cite{RM1966,paper119}
\begin{align}
r_{\rm m}=\frac{1-\eps_{\rm s}(Q,\omega)}{1+
\eps_{\rm s}(Q,\omega)}, \nonumber
\end{align}
where
\begin{align}
\eps_{\rm s}(Q,\omega) = \frac{2 Q }{\pi} \int_0^\infty \frac{dk}{Q^2+k^2}\;\frac{1}{\epsilon_{\rm m}\left(\sqrt{k^2+Q^2},\omega\right)}
\nonumber
\end{align}
plays the role of a surface response function. We further decompose the bulk metal permittivity as \cite{paper119} $\epsilon_{\rm m}(q,\omega)=\epsilon_{\rm b}(\omega)+\epsilon_{\rm free}(q,\omega)-1$, where $\epsilon_{\rm b}$ is the local background response defined above (see Fig.\ S9 in the Appendix) and $\epsilon_{\rm free}$ describes the nonlocal contribution of free conduction electrons. We adopt three different levels of approximation for the latter in the calculations presented in Fig.\ \ref{Fig4}:
\begin{widetext}
\begin{align}
\eps_{\rm hydro}(q,\omega)&=1+
\dfrac{\omega_{\rm p}^2}{\hbar^2\beta^2 q^2 / m_{\rm e}^2-\omega(\omega+\ii \gamma)}, \nonumber\\
\eps_{\rm Lindhard}(q,\omega)&=1+\dfrac{2m_{\rm e} e^2 \kF}{\pi \hbar^2 q^2} \left[1+F(q/\kF,\hbar\omega/\EF)+F(q/\kF,-\hbar\omega/\EF)\right], \nonumber\\
\eps_{\rm Mermin}(q,\omega)& =1+\dfrac{(\omega + \ii \gamma) \left[ \eps_{\rm Lindhard}(\omega + \ii \gamma,q)-1\right]}{\omega + \ii \gamma\left[\eps_{\rm Lindhard}(\omega + \ii \gamma,q)-1 \right]/\left[\eps_{\rm Lindhard}(0,q)-1 \right]}, \nonumber
\end{align}
\end{widetext}
where the subscripts refer to the hydrodynamic \cite{B1933,R1957}, Lindhard \cite{L1954,HL1970}, and Mermin \cite{M1970} models (see main text). Here, $\omega_{\rm p}$ is the bulk plasma frequency of the metal, $\EF=\hbar^2 \kF^2/2 m^*$ is the Fermi energy,
$\kF=(9\pi/4)^{1/3}(1/r_s)$ is the Fermi wave vector expressed in terms of the one-electron radius $r_s$, and we use the function
\begin{align}
F(x,y)= \frac{1}{2x}\left[1-\left(\frac{x^2+y}{2x}\right)^2 \right] \,\log\left( \frac{x^2+2x+y}{x^2-2x+y}\right).
\nonumber
\end{align}
Incidentally, the reflection coefficient derived from the SRM combined with the choice $\eps_{\rm free}=\eps_{\rm hydro}$ coincides with the solution of the classical hydrodynamic equations \cite{paper149}, where the $\beta=\sqrt{3/5}\,\kF$ term accounts for the hydrodynamic pressure. We apply these models to gold and silver by taking $m^*=\me$ and $r_s=3\times$Bohr-radius (i.e., $r_s=0.16\,$nm), as well as values for $\wp$ and $\gamma$ as specified above.

\section{Reflection and transmission coefficients of monolayer graphene}
We adopt the well-known expressions for the reflection and transmission coefficients of a zero-thickness 2D graphene monolayer in the quasistatic limit \cite{paper235}
\begin{align}
r^{\rm 2D}_{\rm g}&=\frac{1}{1-\ii\omega/(2\pi Q\sigma)}, \nonumber\\
t^{\rm 2D}_{\rm g}&=1-r^{\rm 2D}_{\rm g}, \nonumber
\end{align}
where $\sigma(Q,\omega)$ is the wave-vector- and frequency-dependent nonlocal conductivity of graphene, which we evaluate in the RPA \cite{WSS06,HS07}, further introducing an inelastic lifetime $\tau=500$\,fs via Mermin's prescription \cite{M1970}. Incidentally, applying Eq.\ (\ref{rij}) to the above graphene transmission and reflection coefficients, combined with the coefficients $t_{\rm s}=2\sqrt{\epsilon_{\rm d}}/(\epsilon_{\rm d}+1)$ and $r_{\rm s}=(\epsilon_{\rm d}-1)/(\epsilon_{\rm d}+1)$ for the planar surface of a dielectric of permittivity $\epsilon_{\rm d}$, we readily obtain the coefficient $r=-r_{\rm s}+t_{\rm s}^2r^{\rm 2D}_{\rm g}/(1-r_{\rm s}r^{\rm 2D}_{\rm g})=(1-\epsilon_{\rm d}+\xi)/(1+\epsilon_{\rm d}+\xi)$ with $\xi=4\pi\ii\sigma Q/\omega$ for internal reflection from graphene supported by the dielectric, used in the tutorial model presented in the discussion of Fig.\ \ref{Fig1}. In order to account for the finite extension of the carbon 2p orbitals $\varphi_{\rm 2p}(\rb)$ outward from the plane of the graphene monolayer, we introduce effective graphene reflection and transmission coefficients $r_{\rm g}=r^{\rm 2D}_{\rm g} C_{\rm g}^2 e^{-Q d_{\rm g}}$ and $t_{\rm g}=t^{\rm 2D}_{\rm g} C_{\rm g}^2 e^{-Q d_{\rm g}}$, where $d_{\rm g}=0.33$\,nm is the interlayer spacing of graphite and $C_{\rm g}$ is a coupling factor defined as
\begin{align}
C_{\rm g} = \int d^3\rb \, \varphi^2_{\rm 2p}(\rb) e^{-Q z}.
\nonumber
\end{align}
We approximate $C_{\rm g}$ by using a tabulated 2p orbital for an isolated carbon atom \cite{CR1974}. More precisely, $\varphi_{\rm 2p}(\rb) = z \sum_{j=1}^4 \beta_j e^{-\alpha_j r}$, where $\alpha_1=1.10539$,  $\alpha_2=0.61830$,  $\alpha_3=2.26857$, $\alpha_4=5.23303$, $\beta_1=0.4610$, $\beta_2=0.0134$, $\beta_3=1.5905$, and $\beta_4=0.7291$, all in atomic units. The result is plotted in Fig.\ \ref{S10}, where we find $C_{\rm g}=1$ at $Q=0$ in virtue of orbital normalization. The assumption of an effective thickness implies that classically there is always a finite separation $d_{\rm g}/2$ between the carbon nuclei plane in graphene and the surrounding media.

\section{Optical response of a hBN film}
Motivated by recent experimental studies, we present in the Appendix calculations for systems in which the MG and the metal film are separated by a thin layer of hBN (see Figs.\ \ref{S4} and \ref{S5}). The hBN region is taken to have a thickness $d_{\rm d}$ corresponding to an integer number of atomic-layer spacings along the out-of-plane c(1111) direction. We describe this layer through a local anisotropic permittivity with tensor components \cite{GPR1966}
\begin{align}
\eps_i(\omega) = \eps^{\infty}_i + \sum^2_{j=1} \frac{s_{i,j}^2}{\omega_{i,j}^2+\omega(\omega+\ii \gamma_{i,j})}
\nonumber
\end{align}
for parallel ($i=x,y$) or perpendicular ($i=z$) directions relative to the atomic layers. Here, $j=1,2$ runs over oscillators (Lorentzians) with resonance energies $\omega_{x,1}=170.0$, $\omega_{z,1}=97.1$, $\omega_{x,2}=95.1$, $\omega_{z,2}=187.0$; transition strengths $s_{x,1}=232.0$, $s_{z,1}=70.8$, $s_{x,2}=43.5$, $s_{z,2}=126.0$; and dampings $\gamma_{x,1}=3.60$, $\gamma_{z,1}=0.99$, $\gamma_{x,2}=3.40$, $\gamma_{z,2}=9.92$ (all of them in meV). Following the methods of Ref.\ \cite{paper295}, we readily obtain the reflection and transmission coefficients
\begin{align}
r_{\rm d} &=(\eps^2-1)\,\frac{1-\ee^{-2qd_{\rm d}}}{(\eps+1)^2-(\eps-1)^2\,\ee^{-2qd_{\rm d}}}, \nonumber \\
t_{\rm d} &=\frac{4\eps\,\ee^{-qd_{\rm d}}}{(\eps+1)^2-(\eps-1)^2\,\ee^{-2qd_{\rm d}}}, \nonumber
\end{align}
where $\eps = \sqrt{\eps_x\eps_z}$ (with ${\rm Im}\{\eps\}>0$) and $q=Q\sqrt{\eps_x/\eps_z}$ (with ${\rm Re}\{q\}>0$). In the calculations presented in the Appendix, we assume a thickness $d_{\rm d}=1$\,nm, corresponding to 3 MLs of hBN \cite{GCW13}.

\section{\bf Atomic layer potential (ALP)}
In the ALP model we use the parametrized potential of Ref.\ \cite{CSE99} to obtain the one-electron states of a metal film including {\it ad hoc} band-structure information. Specifically, this potential consists of a harmonic bulk component, a differentiated region describing each outermost layer, and a long-range image tail, constructed in such a way that it reproduces several experimentally observed electronic structure features, namely: the work function, the surface-projected bulk gap, and the position of the surface states relative to the Fermi level, all of which depend on material and crystallographic orientation. For simplicity and to a good approximation, we take the effective electron mass as $m^*=m_{\rm e}$ in all directions. For a semi-infinite metal placed in the $z>0$ region, the potential referred to the vacuum level can be written as \cite{CSE99}
\begin{align}
V_{\rm surf}(z) = \left\{
  \begin{array}{l l}
    \frac{1}{4(z-z_{\rm i})}\left[\ee^{-\lambda(z-z_{\rm i})}-1\right], \quad\quad & z<z_{\rm i}\\
    A_3\,\ee^{-\alpha(z-z_{\rm t})}, \quad\quad & z_{\rm i}<z<z_{\rm t}\\
    -A_{20}+A_2\cos{(\eta z)}, \quad\quad & z_{\rm t}<z<0\\
    A_{10}+A_1\cos{(2\pi z /a_s)}, \quad\quad & z>0
  \end{array} \right.
\label{rect}
\end{align} 
where the normal coordinate $z$ is given relative to the position of the outermost atomic plane ($z=0$); $a_s$ is the inter-atomic layer spacing; the coefficients $A_1$ and $A_{10}$ are chosen to reproduce the width and position of the noted energy gap, respectively; the space between $z=z_{\rm i}$ and $z=0$ represents the transition from the solid bulk to free-space, where electron spill out takes place; and the parameters $A_{2}$ and $\eta$ determine the positions of the Fermi level and the surface states relative to the vacuum level. We list the values of these parameters for Au(111) and Ag(111) in Table\ \ref{chulkov_param}. Five of the remaining six parameters are determined by imposing the continuity of the potential and its first derivative, so that $A_{20}=A_2-A_1-A_{10}$, $A_3=-A_{20}+A_2 \cos{(\eta z_{\rm t})}$, $\alpha= \eta A_2 \sin{(\eta z_{\rm t})} / A_3$, $\lambda=2\alpha$, and $z_{\rm i}=\alpha^{-1}\log{\left(-\lambda/ 4 A_3\right)}+z_{\rm t}$, while the intermediate point $z_{\rm t}=-5\pi / 4\eta$ is fixed with respect to the surface atomic layer in such a way that $z_{\rm i}$ corresponds to the image plane position, which is important for describing image states \cite{CSE99}. The potential for a film with outermost atomic planes at $z=a_s/2$ and $z=d_{\rm m}-a_s/2$ can be expressed using Eq.\ (\ref{rect}) as
\begin{align}
V(z) = \left\{ 
  \begin{array}{l l}
    V_{\rm surf}(z-a_s/2), \quad\quad & z<\dm/2\\
    V_{\rm surf}(\dm-a_s/2-z), \quad\quad & z>\dm/2
  \end{array} \right.
\nonumber
\end{align}
which is continuous at $z=d_{\rm m}/2$ by construction. Obviously, $d_{\rm m}$ must be taken to be a multiple of the atomic-layer spacing $a_s$.

\begin{table*} 
\centering
\begin{tabular}{c|c|c|c|c|c}  \hline
 & $a_s$ (nm) & $A_{10}$ (eV) & $A_1$ (eV) & $A_2$ (eV) & $\eta$ (nm$^{-1}$) \\ \hline
Au(111) &  0.2356 & -11.030 & 4.60 & 4.8576 & 53.364 \\ 
Ag(111) &  0.2361 & -9.640  & 4.30 & 3.8442 & 48.470 \\ \hline 
\end{tabular}
\caption{{\bf Atomic layer parameters.} We list the parameters needed to feed Eq.\ (\ref{rect}) in order to describe the ALP for Au(111) and Ag(111) surfaces, taken from Ref.\ \cite{CSE99}.}
\label{chulkov_param}
\end{table*}

\clearpage
\section{Additional figures}

\begin{figure*}[h]
\center
\includegraphics[width=0.77\textwidth]{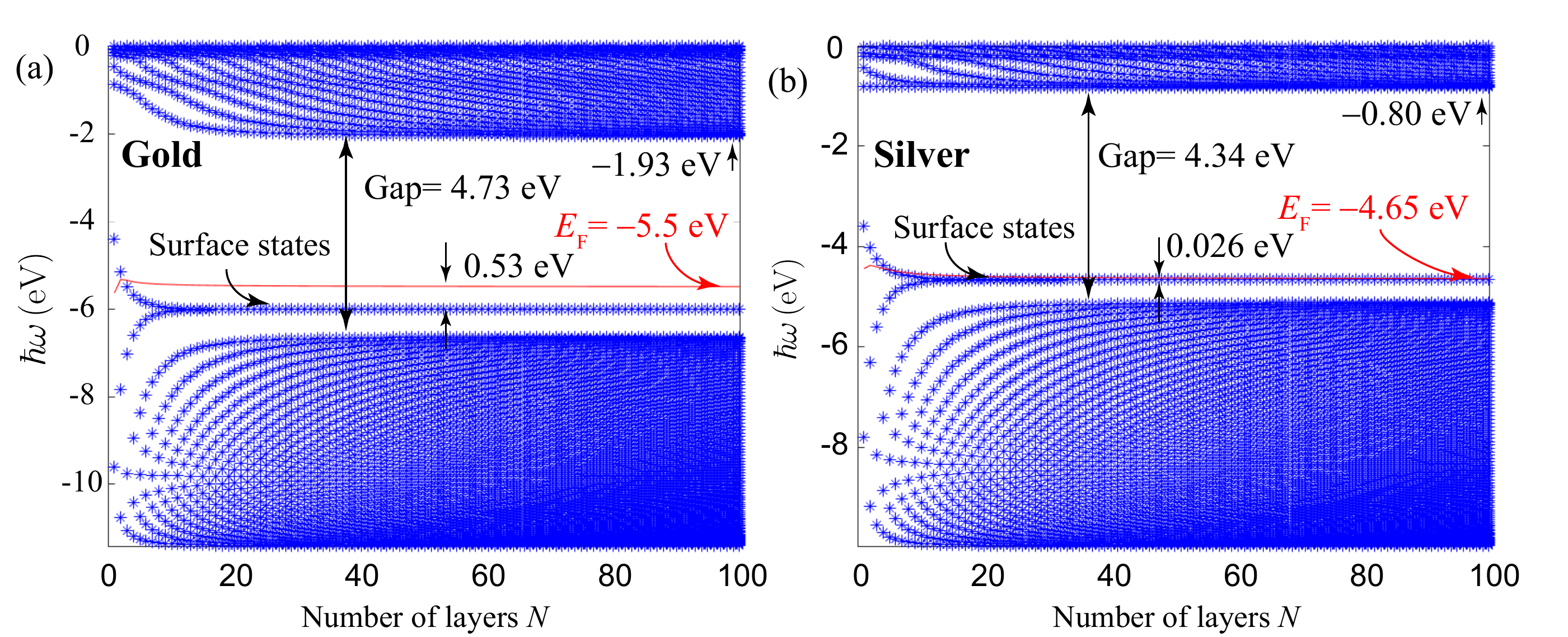}
\caption{\textbf{Vertical electronic states of Au and Ag films as a function of thickness.} Each state with zero parallel wave vector is represented by a symbol as a function of the number of atomic layers $N$. The dependence on parallel wave vector comes through a parabola for each of these states (not shown). The large-$N$-limit energies shown by labels in the plots (the Fermi energy $E_{\rm F}$ relative to vacuum, the gap energy, and the distance from $E_{\rm F}$ to the surface-state energy) reproduce the experiments in Refs.\ \cite{KG1987,ZS1987,PMM95}.}
\label{S1}
\end{figure*}

\begin{figure*}[h]
\center
\includegraphics[width=1\textwidth]{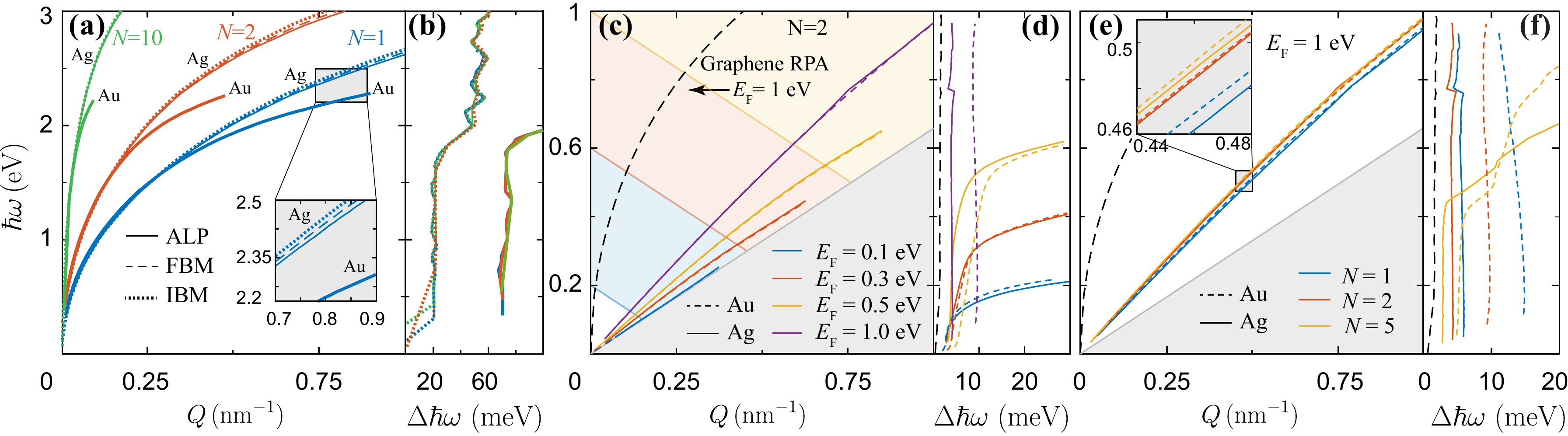}
\caption{\textbf{Comparison of plasmons in monolayer graphene on top of atomically-thin films of either silver or gold.} {\bf (a,b)} Dispersion (a) and FWHM (b) of high-energy plasmons in films comprising $N=1$-10 atomic Ag(111) layers (0.236\,nm thickness per layer) obtained using the RPA. Nearly indistinguishable results are obtained by using the ALP, FBM, and IBM potentials. Results obtained from the ALP for Au (taken from Fig.\ \ref{Fig2}b,c) are shown for comparison (see labels).  {\bf (c-f)} Dispersion relation (c,e) and FWHM (d,f) of acoustic plasmons in graphene-metal films containing $N$ (111) atomic layers of gold (dashed curves) and silver (solid curves) for $N=2$ and various graphene doping levels (c,d), and for $\EF=1\,$eV and various metal thicknesses (e,f). In (c-f) we describe graphene in the RPA and the metal using the ALP model. Results for self-standing graphene with $\EF=1\,$eV (long-dashed curves) are shown for reference.}
\label{S2}
\end{figure*}

\begin{figure*}[h]
\centering
\includegraphics[width=0.65\textwidth]{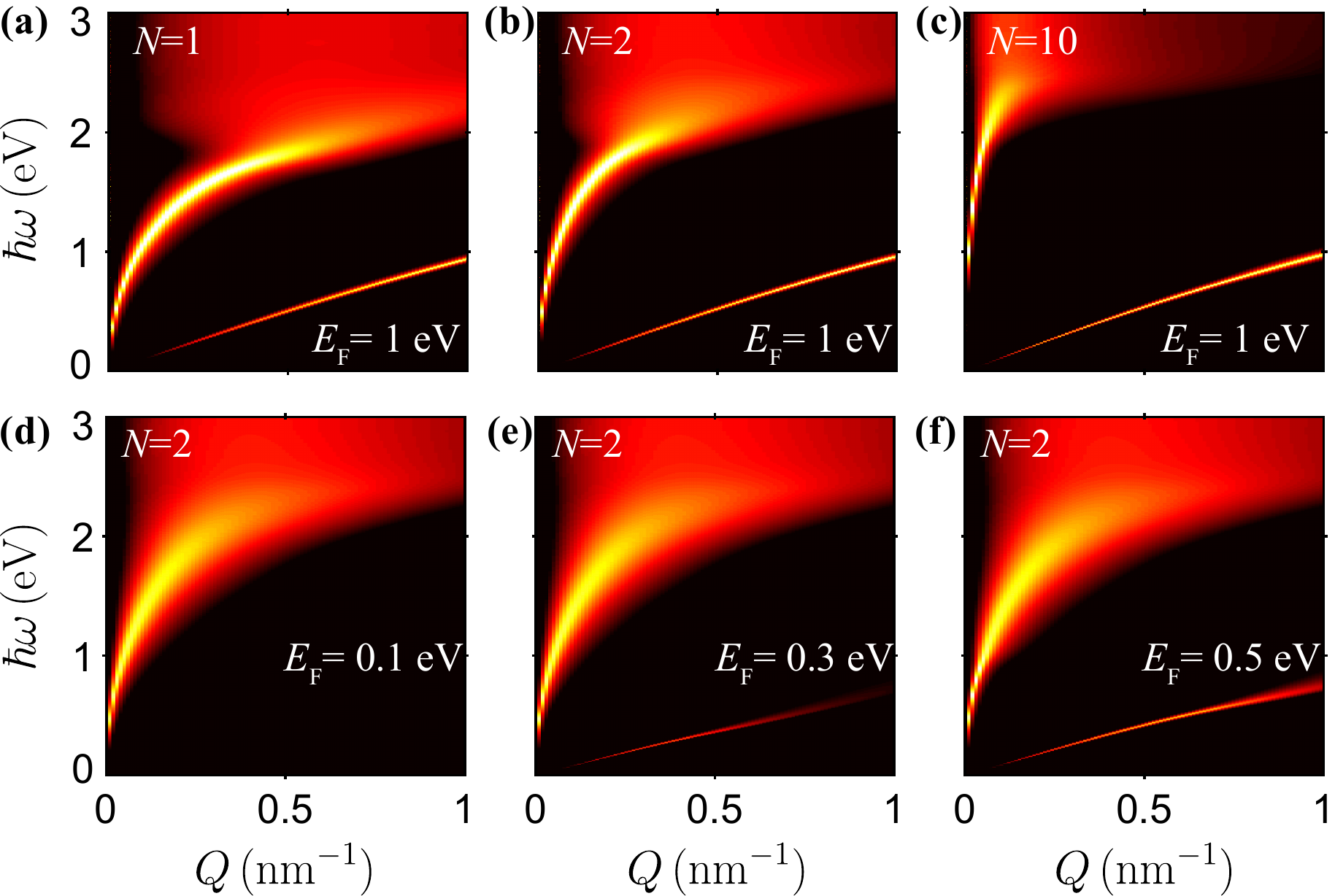}
\caption{\textbf{Dispersion relation of graphene deposited on atomically-thin gold films.} We plot the loss function ${\rm Im}\{R\}$, where $R$ is the reflection coefficient (see Appendix), from which we extract the dispersion relations shown in Fig.\ \ref{Fig3}b (upper plots, varying thickness $N$ for fixed graphene doping $\EF=1\,$eV) and Fig.\ \ref{Fig3}d (lower plots, fixed thickness $N=2$ for varying graphene doping $\EF$).}
\label{S3}
\end{figure*}

\begin{figure*}[h]
\includegraphics[width=1\textwidth]{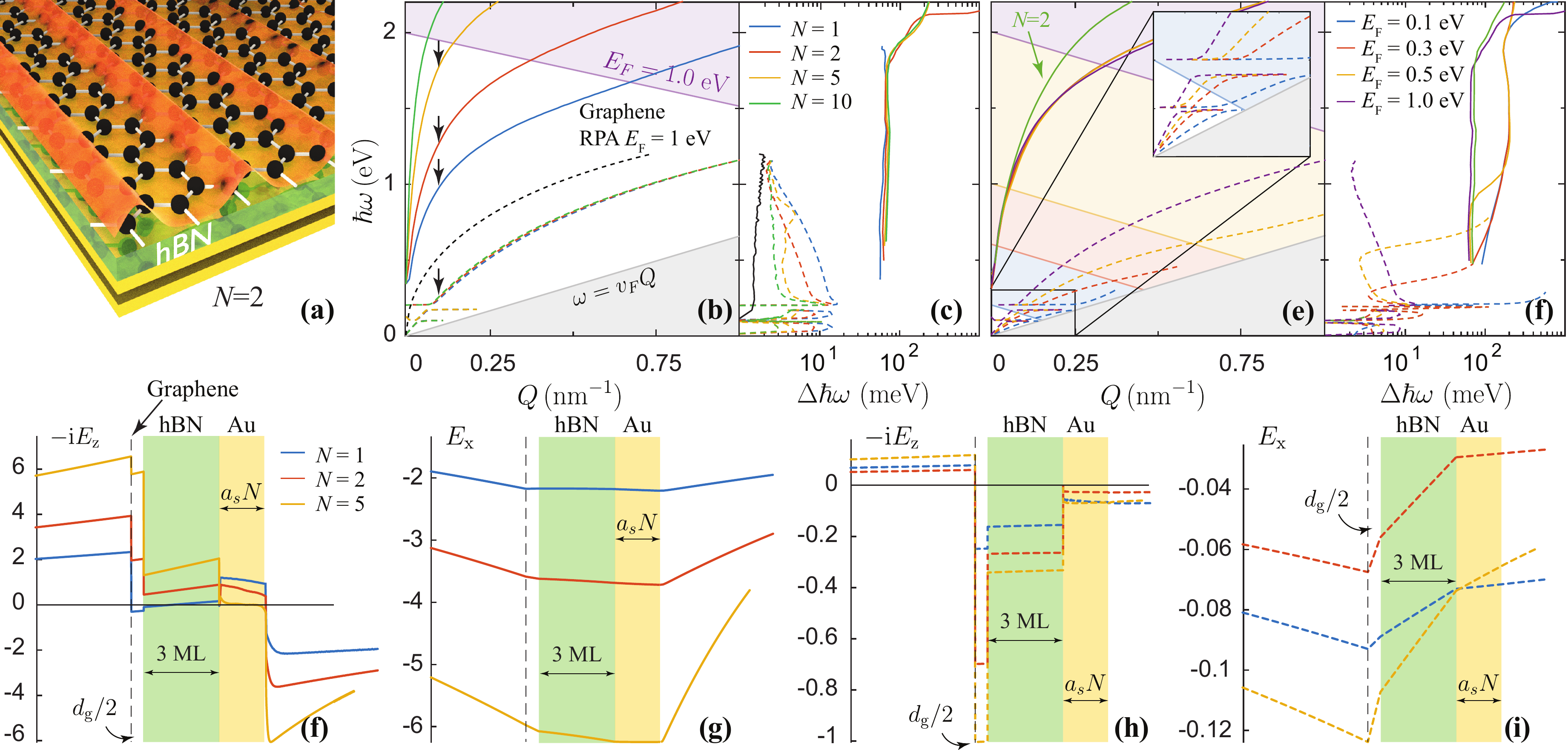}
\caption{\textbf{Thickness and doping dependence of plasmons in graphene-BN-gold film heterostructures.} Same as Fig.\ \ref{Fig3} of the main text with a layer of 1\,nm of hBN (i.e., 3 MLs) separating the graphene monolayer from the metal.}
\label{S4}
\end{figure*}

\begin{figure*}[h]
\includegraphics[width=1\textwidth]{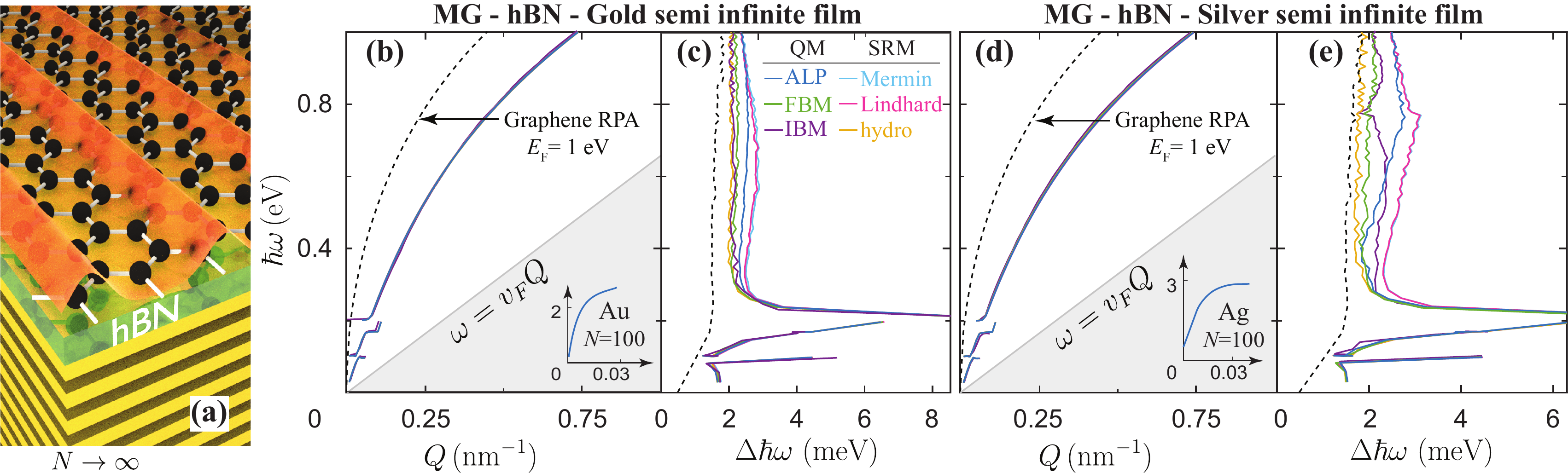}
\caption{\textbf{Model dependence of acoustic plasmons in graphene-BN-metal hybrid films.} Same as Fig.\ \ref{Fig4} of the main text with a layer of 1\,nm of hBN (i.e., 3 MLs) separating the graphene monolayer from the metal.}
\label{S5}
\end{figure*}

\begin{figure*}[h]
\centering
\includegraphics[width=0.65\textwidth]{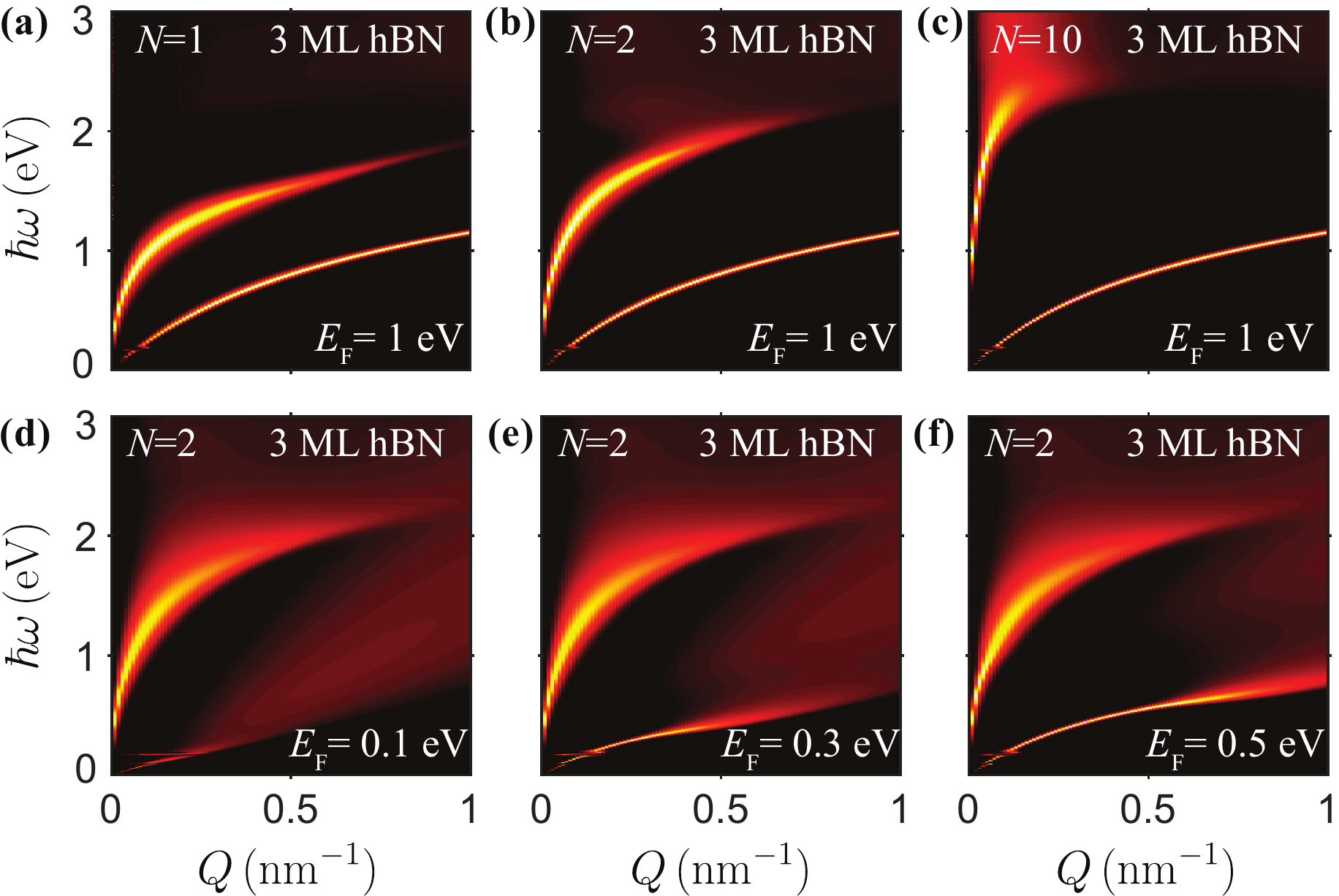}
\caption{\textbf{Dispersion relation of graphene separated by hBN from atomically-thin gold films.} We plot the loss function ${\rm Im}\{R\}$ from which we extract the dispersion relations shown in Fig.\ \ref{S4}b (upper plots, varying metal thickness $N$ for fixed graphene doping $\EF=1\,$eV) and Fig.\ \ref{S4}d (lower plots, fixed metal thickness $N=2$ for varying graphene doping $\EF$). The intermediate hBN layer is 1\,nm thick (i.e., approximately 3\,ML).}
\label{S6}
\end{figure*}

\begin{figure*}[h]
\centering
\includegraphics[width=0.9\textwidth]{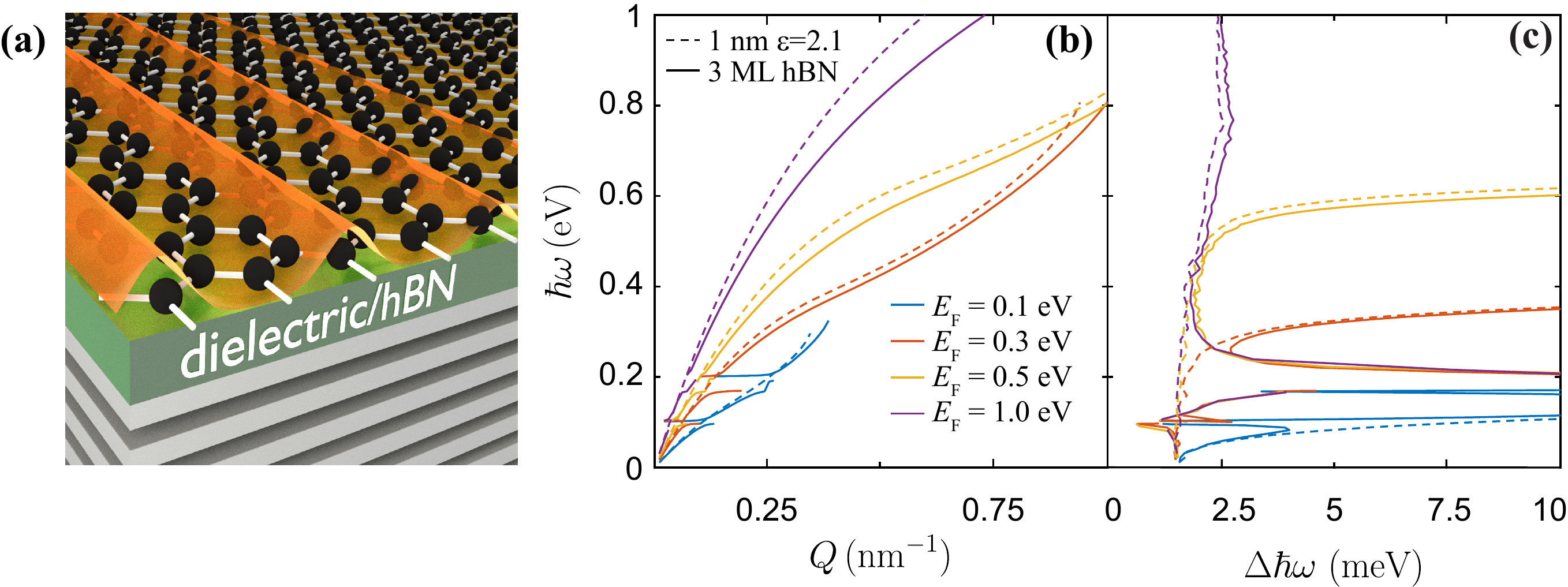}
\caption{\textbf{Doping dependence of acoustic plasmons in a graphene-hBN-semi-inifinite silver structures.} {\bf (a)} Sketch of the system under consideration. The intermediate layer separating the graphene from the metal is either an isotropic dielectric ($\epsilon=2.1$) or hBN, with a thickness of 1\,nm in both cases. {\bf (b,c)} Dispersion (b) and FWHM (c) for different graphene Fermi energies (see legend in (b)) with either an intermediate $\epsilon=2.1$ dielectric (dashed curves) of hBN (solid curves).}
\label{S7}
\end{figure*}

\begin{figure*}[h]
\centering
\includegraphics[width=0.7\textwidth]{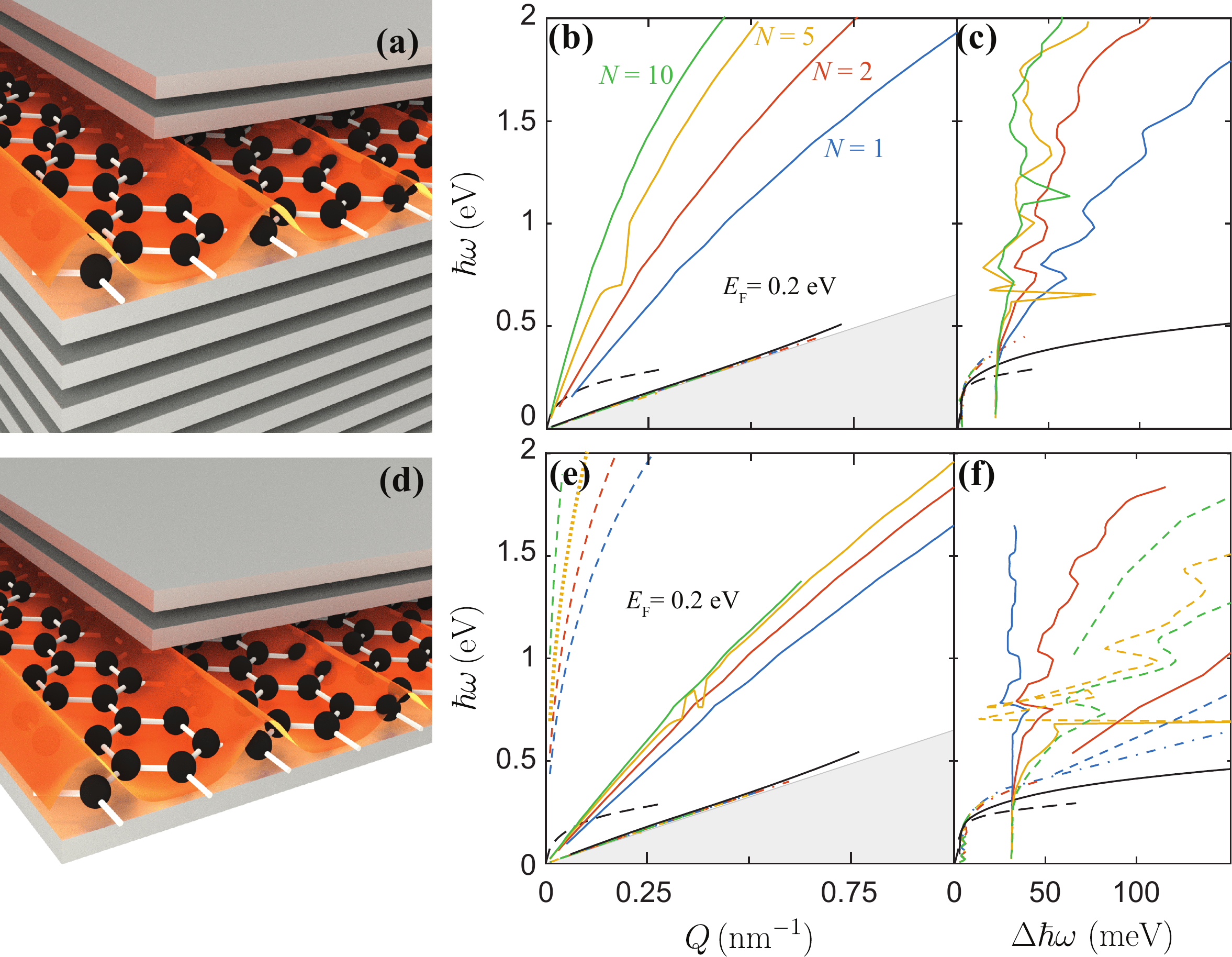}
\caption{\textbf{Plasmons in metal-graphene-metal structures.} Same as Fig.\ \ref{Fig5} of the main text with graphene doping $\EF=0.2\,$eV.}
\label{S8}
\end{figure*}

\begin{figure*}[h]
\centering
\includegraphics[width=0.7\textwidth]{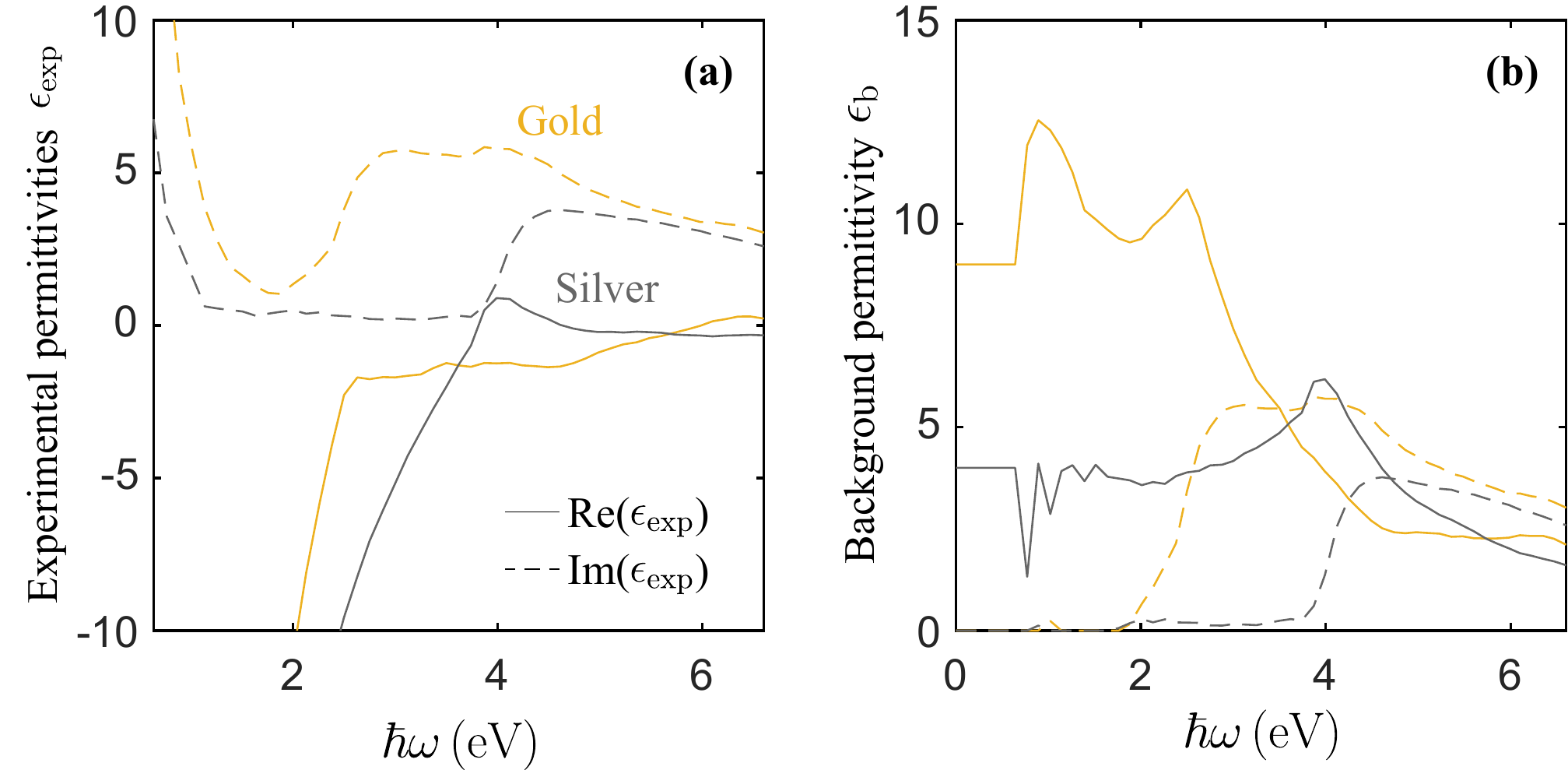}
\caption{\textbf{Dielectric function of gold and silver.} We reproduce in (a) the experimentally measured dielectric function $\epsilon_{\rm exp}(\omega)$ of Au and Ag from Ref.\ \cite{JC1972} (extrapolated as a Drude tail below 0.64\,eV). In (b) we plot the background obtained by removing a Drude contribution (i.e., $\epsilon_{\rm b}(\omega)=\epsilon_{\rm exp}(\omega)+\omega_{\rm p}^2/\omega(\omega+\ii\gamma)$), with parameters $\hbar\omega_{\rm p}=9.06\,$eV, $\hbar\gamma=0.071\,$eV for Au and $\hbar\omega_{\rm p}=9.17\,$eV, $\hbar\gamma=0.021\,$eV for Ag.}
\label{S9}
\end{figure*}

\begin{figure*}[h]
\centering
\includegraphics[width=0.45\textwidth]{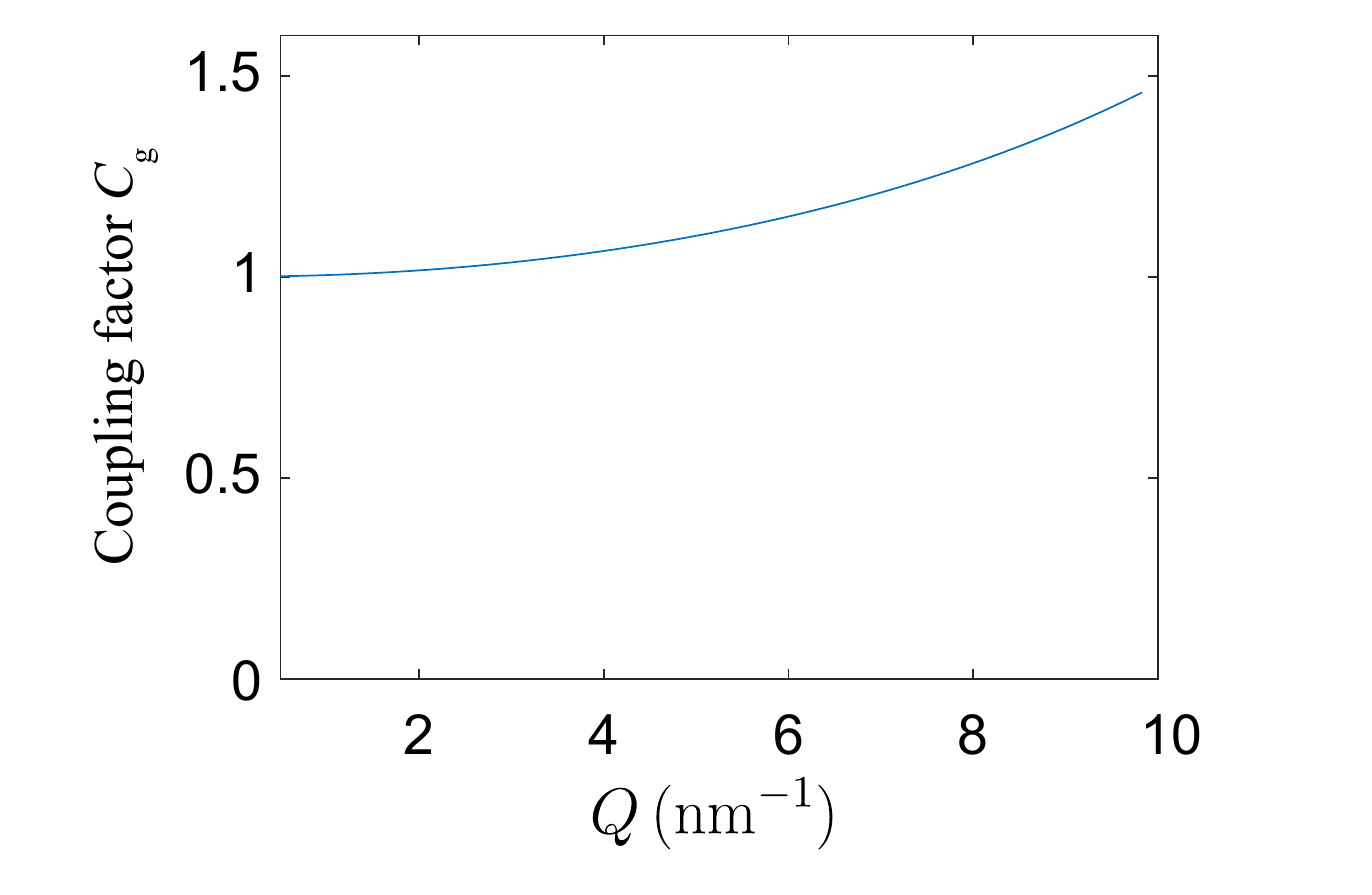}
\caption{\textbf{Coupling factor $C_g$.} We plot the coupling factor that corrects for the finite thickness of graphene as a function of parallel wave vector $Q$ (see main text).}
\label{S10}
\end{figure*}

Next, we present additional Figs.\ \ref{S1}-\ref{S10} referenced in the main text.

\clearpage

\begin{thebibliography}{75}
\expandafter\ifx\csname natexlab\endcsname\relax\def\natexlab#1{#1}\fi
\expandafter\ifx\csname bibnamefont\endcsname\relax
  \def\bibnamefont#1{#1}\fi
\expandafter\ifx\csname bibfnamefont\endcsname\relax
  \def\bibfnamefont#1{#1}\fi
\expandafter\ifx\csname citenamefont\endcsname\relax
  \def\citenamefont#1{#1}\fi
\expandafter\ifx\csname url\endcsname\relax
  \def\url#1{\texttt{#1}}\fi
\expandafter\ifx\csname urlprefix\endcsname\relax\def\urlprefix{URL }\fi
\providecommand{\bibinfo}[2]{#2}
\providecommand{\eprint}[2][]{\url{#2}}

\bibitem[{\citenamefont{Novoselov et~al.}(2004)\citenamefont{Novoselov, Geim,
  Morozov, Jiang, Zhang, Dubonos, Grigorieva, and Firsov}}]{NGM04}
\bibinfo{author}{\bibfnamefont{K.~S.} \bibnamefont{Novoselov}},
  \bibinfo{author}{\bibfnamefont{A.~K.} \bibnamefont{Geim}},
  \bibinfo{author}{\bibfnamefont{S.~V.} \bibnamefont{Morozov}},
  \bibinfo{author}{\bibfnamefont{D.}~\bibnamefont{Jiang}},
  \bibinfo{author}{\bibfnamefont{Y.}~\bibnamefont{Zhang}},
  \bibinfo{author}{\bibfnamefont{S.~V.} \bibnamefont{Dubonos}},
  \bibinfo{author}{\bibfnamefont{I.~V.} \bibnamefont{Grigorieva}},
  \bibnamefont{and} \bibinfo{author}{\bibfnamefont{A.~A.}
  \bibnamefont{Firsov}}, \bibinfo{journal}{Science}
  \textbf{\bibinfo{volume}{306}}, \bibinfo{pages}{666} (\bibinfo{year}{2004}).

\bibitem[{\citenamefont{Xia et~al.}(2014)\citenamefont{Xia, Wang, Xiao, Dubey,
  and Ramasubramaniam}}]{XWX14}
\bibinfo{author}{\bibfnamefont{F.}~\bibnamefont{Xia}},
  \bibinfo{author}{\bibfnamefont{H.}~\bibnamefont{Wang}},
  \bibinfo{author}{\bibfnamefont{D.}~\bibnamefont{Xiao}},
  \bibinfo{author}{\bibfnamefont{M.}~\bibnamefont{Dubey}}, \bibnamefont{and}
  \bibinfo{author}{\bibfnamefont{A.}~\bibnamefont{Ramasubramaniam}},
  \bibinfo{journal}{Nat.\ Photon.} \textbf{\bibinfo{volume}{8}},
  \bibinfo{pages}{899} (\bibinfo{year}{2014}).

\bibitem[{\citenamefont{{Alcaraz Iranzo} et~al.}(2018)\citenamefont{{Alcaraz
  Iranzo}, Nanot, Dias, Epstein, Peng, Efetov, Lundeberg, Parret, Osmond, Hong
  et~al.}}]{AND18}
\bibinfo{author}{\bibfnamefont{D.}~\bibnamefont{{Alcaraz Iranzo}}},
  \bibinfo{author}{\bibfnamefont{S.}~\bibnamefont{Nanot}},
  \bibinfo{author}{\bibfnamefont{E.~J.~C.} \bibnamefont{Dias}},
  \bibinfo{author}{\bibfnamefont{I.}~\bibnamefont{Epstein}},
  \bibinfo{author}{\bibfnamefont{C.}~\bibnamefont{Peng}},
  \bibinfo{author}{\bibfnamefont{D.~K.} \bibnamefont{Efetov}},
  \bibinfo{author}{\bibfnamefont{M.~B.} \bibnamefont{Lundeberg}},
  \bibinfo{author}{\bibfnamefont{R.}~\bibnamefont{Parret}},
  \bibinfo{author}{\bibfnamefont{J.}~\bibnamefont{Osmond}},
  \bibinfo{author}{\bibfnamefont{J.-Y.} \bibnamefont{Hong}},
  \bibnamefont{et~al.}, \bibinfo{journal}{Science}
  \textbf{\bibinfo{volume}{360}}, \bibinfo{pages}{291} (\bibinfo{year}{2018}).

\bibitem[{\citenamefont{Cox and {Garc\'{\i}a de Abajo}}(2014)}]{paper247}
\bibinfo{author}{\bibfnamefont{J.~D.} \bibnamefont{Cox}} \bibnamefont{and}
  \bibinfo{author}{\bibfnamefont{F.~J.} \bibnamefont{{Garc\'{\i}a de Abajo}}},
  \bibinfo{journal}{Nat.\ Commun.} \textbf{\bibinfo{volume}{5}},
  \bibinfo{pages}{5725} (\bibinfo{year}{2014}).

\bibitem[{\citenamefont{Lee and El-Sayed}(2006)}]{KM06}
\bibinfo{author}{\bibfnamefont{K.-S.} \bibnamefont{Lee}} \bibnamefont{and}
  \bibinfo{author}{\bibfnamefont{M.~A.} \bibnamefont{El-Sayed}},
  \bibinfo{journal}{J.\ Phys.\ Chem.\ B} \textbf{\bibinfo{volume}{110}},
  \bibinfo{pages}{19220} (\bibinfo{year}{2006}).

\bibitem[{\citenamefont{Geim and Grigorieva}(2013)}]{GG13}
\bibinfo{author}{\bibfnamefont{A.~K.} \bibnamefont{Geim}} \bibnamefont{and}
  \bibinfo{author}{\bibfnamefont{I.~V.} \bibnamefont{Grigorieva}},
  \bibinfo{journal}{Nature} p. \bibinfo{pages}{419} (\bibinfo{year}{2013}).

\bibitem[{\citenamefont{Basov et~al.}(2016)\citenamefont{Basov, Fogler, and
  {Garc\'{\i}a de Abajo}}}]{paper283}
\bibinfo{author}{\bibfnamefont{D.~N.} \bibnamefont{Basov}},
  \bibinfo{author}{\bibfnamefont{M.~M.} \bibnamefont{Fogler}},
  \bibnamefont{and} \bibinfo{author}{\bibfnamefont{F.~J.}
  \bibnamefont{{Garc\'{\i}a de Abajo}}}, \bibinfo{journal}{Science}
  \textbf{\bibinfo{volume}{354}}, \bibinfo{pages}{aag1992}
  (\bibinfo{year}{2016}).

\bibitem[{\citenamefont{Economou}(1969)}]{E1969}
\bibinfo{author}{\bibfnamefont{E.~N.} \bibnamefont{Economou}},
  \bibinfo{journal}{Phys.\ Rev.} \textbf{\bibinfo{volume}{182}},
  \bibinfo{pages}{539} (\bibinfo{year}{1969}).

\bibitem[{\citenamefont{Barnes et~al.}(2003)\citenamefont{Barnes, Dereux, and
  Ebbesen}}]{BDE03}
\bibinfo{author}{\bibfnamefont{W.~L.} \bibnamefont{Barnes}},
  \bibinfo{author}{\bibfnamefont{A.}~\bibnamefont{Dereux}}, \bibnamefont{and}
  \bibinfo{author}{\bibfnamefont{T.~W.} \bibnamefont{Ebbesen}},
  \bibinfo{journal}{Nature} \textbf{\bibinfo{volume}{424}},
  \bibinfo{pages}{824} (\bibinfo{year}{2003}).

\bibitem[{\citenamefont{Mulvaney et~al.}(2006)\citenamefont{Mulvaney,
  P\'{e}rez-Juste, Giersig, Liz-Marz\'{a}n, and Pecharrom\'{a}n}}]{MPG06}
\bibinfo{author}{\bibfnamefont{P.}~\bibnamefont{Mulvaney}},
  \bibinfo{author}{\bibfnamefont{J.}~\bibnamefont{P\'{e}rez-Juste}},
  \bibinfo{author}{\bibfnamefont{M.}~\bibnamefont{Giersig}},
  \bibinfo{author}{\bibfnamefont{L.~M.} \bibnamefont{Liz-Marz\'{a}n}},
  \bibnamefont{and}
  \bibinfo{author}{\bibfnamefont{C.}~\bibnamefont{Pecharrom\'{a}n}},
  \bibinfo{journal}{Plasmonics} \textbf{\bibinfo{volume}{1}},
  \bibinfo{pages}{61} (\bibinfo{year}{2006}).

\bibitem[{\citenamefont{Khurgin}(2015)}]{K15_2}
\bibinfo{author}{\bibfnamefont{J.~B.} \bibnamefont{Khurgin}},
  \bibinfo{journal}{Nat.\ Nanotech.} \textbf{\bibinfo{volume}{10}},
  \bibinfo{pages}{2} (\bibinfo{year}{2015}).

\bibitem[{\citenamefont{Fei et~al.}(2011)\citenamefont{Fei, Andreev, Bao,
  Zhang, McLeod, Wang, Stewart, Zhao, Dominguez, Thiemens et~al.}}]{FAB11}
\bibinfo{author}{\bibfnamefont{Z.}~\bibnamefont{Fei}},
  \bibinfo{author}{\bibfnamefont{G.~O.} \bibnamefont{Andreev}},
  \bibinfo{author}{\bibfnamefont{W.}~\bibnamefont{Bao}},
  \bibinfo{author}{\bibfnamefont{L.~M.} \bibnamefont{Zhang}},
  \bibinfo{author}{\bibfnamefont{A.~S.} \bibnamefont{McLeod}},
  \bibinfo{author}{\bibfnamefont{C.}~\bibnamefont{Wang}},
  \bibinfo{author}{\bibfnamefont{M.~K.} \bibnamefont{Stewart}},
  \bibinfo{author}{\bibfnamefont{Z.}~\bibnamefont{Zhao}},
  \bibinfo{author}{\bibfnamefont{G.}~\bibnamefont{Dominguez}},
  \bibinfo{author}{\bibfnamefont{M.}~\bibnamefont{Thiemens}},
  \bibnamefont{et~al.}, \bibinfo{journal}{Nano\ Lett.}
  \textbf{\bibinfo{volume}{11}}, \bibinfo{pages}{4701} (\bibinfo{year}{2011}).

\bibitem[{\citenamefont{Chen et~al.}(2012)\citenamefont{Chen, Badioli,
  Alonso-Gonz\'alez, Thongrattanasiri, Huth, Osmond, Spasenovi\'c, Centeno,
  Pesquera, Godignon et~al.}}]{paper196}
\bibinfo{author}{\bibfnamefont{J.}~\bibnamefont{Chen}},
  \bibinfo{author}{\bibfnamefont{M.}~\bibnamefont{Badioli}},
  \bibinfo{author}{\bibfnamefont{P.}~\bibnamefont{Alonso-Gonz\'alez}},
  \bibinfo{author}{\bibfnamefont{S.}~\bibnamefont{Thongrattanasiri}},
  \bibinfo{author}{\bibfnamefont{F.}~\bibnamefont{Huth}},
  \bibinfo{author}{\bibfnamefont{J.}~\bibnamefont{Osmond}},
  \bibinfo{author}{\bibfnamefont{M.}~\bibnamefont{Spasenovi\'c}},
  \bibinfo{author}{\bibfnamefont{A.}~\bibnamefont{Centeno}},
  \bibinfo{author}{\bibfnamefont{A.}~\bibnamefont{Pesquera}},
  \bibinfo{author}{\bibfnamefont{P.}~\bibnamefont{Godignon}},
  \bibnamefont{et~al.}, \bibinfo{journal}{Nature}
  \textbf{\bibinfo{volume}{487}}, \bibinfo{pages}{77} (\bibinfo{year}{2012}).

\bibitem[{\citenamefont{Fei et~al.}(2012)\citenamefont{Fei, Rodin, Andreev,
  Bao, McLeod, Wagner, Zhang, Zhao, Thiemens, Dominguez et~al.}}]{FRA12}
\bibinfo{author}{\bibfnamefont{Z.}~\bibnamefont{Fei}},
  \bibinfo{author}{\bibfnamefont{A.~S.} \bibnamefont{Rodin}},
  \bibinfo{author}{\bibfnamefont{G.~O.} \bibnamefont{Andreev}},
  \bibinfo{author}{\bibfnamefont{W.}~\bibnamefont{Bao}},
  \bibinfo{author}{\bibfnamefont{A.~S.} \bibnamefont{McLeod}},
  \bibinfo{author}{\bibfnamefont{M.}~\bibnamefont{Wagner}},
  \bibinfo{author}{\bibfnamefont{L.~M.} \bibnamefont{Zhang}},
  \bibinfo{author}{\bibfnamefont{Z.}~\bibnamefont{Zhao}},
  \bibinfo{author}{\bibfnamefont{M.}~\bibnamefont{Thiemens}},
  \bibinfo{author}{\bibfnamefont{G.}~\bibnamefont{Dominguez}},
  \bibnamefont{et~al.}, \bibinfo{journal}{Nature}
  \textbf{\bibinfo{volume}{487}}, \bibinfo{pages}{82} (\bibinfo{year}{2012}).

\bibitem[{\citenamefont{Woessner et~al.}(2015)\citenamefont{Woessner,
  Lundeberg, Gao, Principi, Alonso-Gonz\'alez, Carrega, Watanabe, Taniguchi,
  Vignale, Polini et~al.}}]{WLG15}
\bibinfo{author}{\bibfnamefont{A.}~\bibnamefont{Woessner}},
  \bibinfo{author}{\bibfnamefont{M.~B.} \bibnamefont{Lundeberg}},
  \bibinfo{author}{\bibfnamefont{Y.}~\bibnamefont{Gao}},
  \bibinfo{author}{\bibfnamefont{A.}~\bibnamefont{Principi}},
  \bibinfo{author}{\bibfnamefont{P.}~\bibnamefont{Alonso-Gonz\'alez}},
  \bibinfo{author}{\bibfnamefont{M.}~\bibnamefont{Carrega}},
  \bibinfo{author}{\bibfnamefont{K.}~\bibnamefont{Watanabe}},
  \bibinfo{author}{\bibfnamefont{T.}~\bibnamefont{Taniguchi}},
  \bibinfo{author}{\bibfnamefont{G.}~\bibnamefont{Vignale}},
  \bibinfo{author}{\bibfnamefont{M.}~\bibnamefont{Polini}},
  \bibnamefont{et~al.}, \bibinfo{journal}{Nat.\ Mater.}
  \textbf{\bibinfo{volume}{14}}, \bibinfo{pages}{421} (\bibinfo{year}{2015}).

\bibitem[{\citenamefont{Ni et~al.}(2018)\citenamefont{Ni, McLeod, Sun, Wang,
  Xiong, Post, Sunku, Jiang, Hone, Dean et~al.}}]{NMS18}
\bibinfo{author}{\bibfnamefont{G.~X.} \bibnamefont{Ni}},
  \bibinfo{author}{\bibfnamefont{A.~S.} \bibnamefont{McLeod}},
  \bibinfo{author}{\bibfnamefont{Z.}~\bibnamefont{Sun}},
  \bibinfo{author}{\bibfnamefont{L.}~\bibnamefont{Wang}},
  \bibinfo{author}{\bibfnamefont{L.}~\bibnamefont{Xiong}},
  \bibinfo{author}{\bibfnamefont{K.~W.} \bibnamefont{Post}},
  \bibinfo{author}{\bibfnamefont{S.~S.} \bibnamefont{Sunku}},
  \bibinfo{author}{\bibfnamefont{B.-Y.} \bibnamefont{Jiang}},
  \bibinfo{author}{\bibfnamefont{J.}~\bibnamefont{Hone}},
  \bibinfo{author}{\bibfnamefont{C.~R.} \bibnamefont{Dean}},
  \bibnamefont{et~al.}, \bibinfo{journal}{Nature}
  \textbf{\bibinfo{volume}{557}}, \bibinfo{pages}{530} (\bibinfo{year}{2018}).

\bibitem[{\citenamefont{{Garc\'{\i}a de Abajo}}(2014)}]{paper235}
\bibinfo{author}{\bibfnamefont{F.~J.} \bibnamefont{{Garc\'{\i}a de Abajo}}},
  \bibinfo{journal}{ACS\ Photon.} \textbf{\bibinfo{volume}{1}},
  \bibinfo{pages}{135} (\bibinfo{year}{2014}).

\bibitem[{\citenamefont{Emani et~al.}(2012)\citenamefont{Emani, Chung, Ni,
  Kildishev, Chen, and Boltasseva}}]{ECN12}
\bibinfo{author}{\bibfnamefont{N.~K.} \bibnamefont{Emani}},
  \bibinfo{author}{\bibfnamefont{T.-F.} \bibnamefont{Chung}},
  \bibinfo{author}{\bibfnamefont{X.}~\bibnamefont{Ni}},
  \bibinfo{author}{\bibfnamefont{A.~V.} \bibnamefont{Kildishev}},
  \bibinfo{author}{\bibfnamefont{Y.~P.} \bibnamefont{Chen}}, \bibnamefont{and}
  \bibinfo{author}{\bibfnamefont{A.}~\bibnamefont{Boltasseva}},
  \bibinfo{journal}{Nano\ Lett.} \textbf{\bibinfo{volume}{12}},
  \bibinfo{pages}{5202} (\bibinfo{year}{2012}).

\bibitem[{\citenamefont{Yu et~al.}(2016)\citenamefont{Yu, Pruneri, and
  {Garc\'{\i}a de Abajo}}}]{paper277}
\bibinfo{author}{\bibfnamefont{R.}~\bibnamefont{Yu}},
  \bibinfo{author}{\bibfnamefont{V.}~\bibnamefont{Pruneri}}, \bibnamefont{and}
  \bibinfo{author}{\bibfnamefont{F.~J.} \bibnamefont{{Garc\'{\i}a de Abajo}}},
  \bibinfo{journal}{Sci.\ Rep.} \textbf{\bibinfo{volume}{6}},
  \bibinfo{pages}{32144} (\bibinfo{year}{2016}).

\bibitem[{\citenamefont{Lundeberg et~al.}(2017)\citenamefont{Lundeberg, Gao,
  Asgari, Tan, Duppen, Autore, Alonso-Gonz\'alez, Woessner, Watanabe, Taniguchi
  et~al.}}]{LGA17}
\bibinfo{author}{\bibfnamefont{M.~B.} \bibnamefont{Lundeberg}},
  \bibinfo{author}{\bibfnamefont{Y.}~\bibnamefont{Gao}},
  \bibinfo{author}{\bibfnamefont{R.}~\bibnamefont{Asgari}},
  \bibinfo{author}{\bibfnamefont{C.}~\bibnamefont{Tan}},
  \bibinfo{author}{\bibfnamefont{B.~V.} \bibnamefont{Duppen}},
  \bibinfo{author}{\bibfnamefont{M.}~\bibnamefont{Autore}},
  \bibinfo{author}{\bibfnamefont{P.}~\bibnamefont{Alonso-Gonz\'alez}},
  \bibinfo{author}{\bibfnamefont{A.}~\bibnamefont{Woessner}},
  \bibinfo{author}{\bibfnamefont{K.}~\bibnamefont{Watanabe}},
  \bibinfo{author}{\bibfnamefont{T.}~\bibnamefont{Taniguchi}},
  \bibnamefont{et~al.}, \bibinfo{journal}{Science}
  \textbf{\bibinfo{volume}{89}}, \bibinfo{pages}{035004}
  (\bibinfo{year}{2017}).

\bibitem[{\citenamefont{Iranzo et~al.}(2018)\citenamefont{Iranzo, Nanot, Dias,
  Epstein, Peng, Efetov, Lundeberg, Parret, Osmond, Hong et~al.}}]{IND18}
\bibinfo{author}{\bibfnamefont{D.~A.} \bibnamefont{Iranzo}},
  \bibinfo{author}{\bibfnamefont{S.}~\bibnamefont{Nanot}},
  \bibinfo{author}{\bibfnamefont{E.~J.~C.} \bibnamefont{Dias}},
  \bibinfo{author}{\bibfnamefont{I.}~\bibnamefont{Epstein}},
  \bibinfo{author}{\bibfnamefont{C.}~\bibnamefont{Peng}},
  \bibinfo{author}{\bibfnamefont{D.~K.} \bibnamefont{Efetov}},
  \bibinfo{author}{\bibfnamefont{M.~B.} \bibnamefont{Lundeberg}},
  \bibinfo{author}{\bibfnamefont{R.}~\bibnamefont{Parret}},
  \bibinfo{author}{\bibfnamefont{J.}~\bibnamefont{Osmond}},
  \bibinfo{author}{\bibfnamefont{J.-Y.} \bibnamefont{Hong}},
  \bibnamefont{et~al.}, \bibinfo{journal}{Science}
  \textbf{\bibinfo{volume}{360}}, \bibinfo{pages}{291} (\bibinfo{year}{2018}).

\bibitem[{\citenamefont{Dionne et~al.}(2006)\citenamefont{Dionne, Sweatlock,
  Atwater, and Polman}}]{DSA06}
\bibinfo{author}{\bibfnamefont{J.~A.} \bibnamefont{Dionne}},
  \bibinfo{author}{\bibfnamefont{L.~A.} \bibnamefont{Sweatlock}},
  \bibinfo{author}{\bibfnamefont{H.~A.} \bibnamefont{Atwater}},
  \bibnamefont{and} \bibinfo{author}{\bibfnamefont{A.}~\bibnamefont{Polman}},
  \bibinfo{journal}{Phys.\ Rev.\ B} \textbf{\bibinfo{volume}{73}},
  \bibinfo{pages}{035407} (\bibinfo{year}{2006}).

\bibitem[{\citenamefont{Principi et~al.}(2011)\citenamefont{Principi, Asgari,
  and Polini}}]{PAP11}
\bibinfo{author}{\bibfnamefont{A.}~\bibnamefont{Principi}},
  \bibinfo{author}{\bibfnamefont{R.}~\bibnamefont{Asgari}}, \bibnamefont{and}
  \bibinfo{author}{\bibfnamefont{M.}~\bibnamefont{Polini}},
  \bibinfo{journal}{Solid\ State\ Commun.} \textbf{\bibinfo{volume}{151}},
  \bibinfo{pages}{1627} (\bibinfo{year}{2011}).

\bibitem[{\citenamefont{Principi et~al.}(2018)\citenamefont{Principi, {van
  Loon}, Polini, and Katsnelson}}]{PVP18}
\bibinfo{author}{\bibfnamefont{A.}~\bibnamefont{Principi}},
  \bibinfo{author}{\bibfnamefont{E.}~\bibnamefont{{van Loon}}},
  \bibinfo{author}{\bibfnamefont{M.}~\bibnamefont{Polini}}, \bibnamefont{and}
  \bibinfo{author}{\bibfnamefont{M.~I.} \bibnamefont{Katsnelson}},
  \bibinfo{journal}{Phys.\ Rev.\ B} \textbf{\bibinfo{volume}{98}},
  \bibinfo{pages}{035427} (\bibinfo{year}{2018}).

\bibitem[{\citenamefont{Dias et~al.}(2018)\citenamefont{Dias, Iranzo,
  Gon\c{c}alves, Hajati, Bludov, Jauho, Mortensen, Koppens, and Peres}}]{DIG18}
\bibinfo{author}{\bibfnamefont{E.~J.~C.} \bibnamefont{Dias}},
  \bibinfo{author}{\bibfnamefont{D.~A.} \bibnamefont{Iranzo}},
  \bibinfo{author}{\bibfnamefont{P.~A.~D.} \bibnamefont{Gon\c{c}alves}},
  \bibinfo{author}{\bibfnamefont{Y.}~\bibnamefont{Hajati}},
  \bibinfo{author}{\bibfnamefont{Y.~V.} \bibnamefont{Bludov}},
  \bibinfo{author}{\bibfnamefont{A.-P.} \bibnamefont{Jauho}},
  \bibinfo{author}{\bibfnamefont{N.~A.} \bibnamefont{Mortensen}},
  \bibinfo{author}{\bibfnamefont{F.~H.~L.} \bibnamefont{Koppens}},
  \bibnamefont{and} \bibinfo{author}{\bibfnamefont{N.~M.~R.}
  \bibnamefont{Peres}}, \bibinfo{journal}{Phys.\ Rev.\ B}
  \textbf{\bibinfo{volume}{97}}, \bibinfo{pages}{245405}
  (\bibinfo{year}{2018}).

\bibitem[{\citenamefont{Bloch}(1933)}]{B1933}
\bibinfo{author}{\bibfnamefont{F.}~\bibnamefont{Bloch}}, \bibinfo{journal}{Z.\
  Phys.} \textbf{\bibinfo{volume}{81}}, \bibinfo{pages}{363}
  (\bibinfo{year}{1933}).

\bibitem[{\citenamefont{Ritchie}(1957)}]{R1957}
\bibinfo{author}{\bibfnamefont{R.~H.} \bibnamefont{Ritchie}},
  \bibinfo{journal}{Phys.\ Rev.} \textbf{\bibinfo{volume}{106}},
  \bibinfo{pages}{874} (\bibinfo{year}{1957}).

\bibitem[{\citenamefont{David and {Garc\'{\i}a de Abajo}}(2014)}]{paper244}
\bibinfo{author}{\bibfnamefont{C.}~\bibnamefont{David}} \bibnamefont{and}
  \bibinfo{author}{\bibfnamefont{F.~J.} \bibnamefont{{Garc\'{\i}a de Abajo}}},
  \bibinfo{journal}{ACS\ Nano} \textbf{\bibinfo{volume}{8}},
  \bibinfo{pages}{9558} (\bibinfo{year}{2014}).

\bibitem[{\citenamefont{Mortensen et~al.}(2014)\citenamefont{Mortensen, Raza,
  Wubs, S{\o}ndergaard, and Bozhevolnyi}}]{MRW14}
\bibinfo{author}{\bibfnamefont{N.~A.} \bibnamefont{Mortensen}},
  \bibinfo{author}{\bibfnamefont{S.}~\bibnamefont{Raza}},
  \bibinfo{author}{\bibfnamefont{M.}~\bibnamefont{Wubs}},
  \bibinfo{author}{\bibfnamefont{T.}~\bibnamefont{S{\o}ndergaard}},
  \bibnamefont{and} \bibinfo{author}{\bibfnamefont{S.~I.}
  \bibnamefont{Bozhevolnyi}}, \bibinfo{journal}{Nat.\ Commun.}
  \textbf{\bibinfo{volume}{5}}, \bibinfo{pages}{3809} (\bibinfo{year}{2014}).

\bibitem[{\citenamefont{Raza et~al.}(2013{\natexlab{a}})\citenamefont{Raza,
  Christensen, Wubs, Bozhevolnyi, and Mortensen}}]{RCW13}
\bibinfo{author}{\bibfnamefont{S.}~\bibnamefont{Raza}},
  \bibinfo{author}{\bibfnamefont{T.}~\bibnamefont{Christensen}},
  \bibinfo{author}{\bibfnamefont{M.}~\bibnamefont{Wubs}},
  \bibinfo{author}{\bibfnamefont{S.~I.} \bibnamefont{Bozhevolnyi}},
  \bibnamefont{and} \bibinfo{author}{\bibfnamefont{N.~A.}
  \bibnamefont{Mortensen}}, \bibinfo{journal}{Phys.\ Rev.\ B}
  \textbf{\bibinfo{volume}{88}}, \bibinfo{pages}{115401}
  (\bibinfo{year}{2013}{\natexlab{a}}).

\bibitem[{\citenamefont{Moreau et~al.}(2013)\citenamefont{Moreau, Cirac\`i, and
  Smith}}]{MCS13}
\bibinfo{author}{\bibfnamefont{A.}~\bibnamefont{Moreau}},
  \bibinfo{author}{\bibfnamefont{C.}~\bibnamefont{Cirac\`i}}, \bibnamefont{and}
  \bibinfo{author}{\bibfnamefont{D.~R.} \bibnamefont{Smith}},
  \bibinfo{journal}{Phys.\ Rev.\ B} \textbf{\bibinfo{volume}{87}},
  \bibinfo{pages}{045401} (\bibinfo{year}{2013}).

\bibitem[{\citenamefont{David et~al.}(2013)\citenamefont{David, Mortensen, and
  Christensen}}]{DMC13}
\bibinfo{author}{\bibfnamefont{C.}~\bibnamefont{David}},
  \bibinfo{author}{\bibfnamefont{N.~A.} \bibnamefont{Mortensen}},
  \bibnamefont{and}
  \bibinfo{author}{\bibfnamefont{J.}~\bibnamefont{Christensen}},
  \bibinfo{journal}{Sci.\ Rep.} \textbf{\bibinfo{volume}{3}},
  \bibinfo{pages}{2526} (\bibinfo{year}{2013}).

\bibitem[{\citenamefont{Ciraci et~al.}(2013)\citenamefont{Ciraci, Pendry, and
  Smith}}]{CPS13}
\bibinfo{author}{\bibfnamefont{C.}~\bibnamefont{Ciraci}},
  \bibinfo{author}{\bibfnamefont{J.~B.} \bibnamefont{Pendry}},
  \bibnamefont{and} \bibinfo{author}{\bibfnamefont{D.~R.} \bibnamefont{Smith}},
  \bibinfo{journal}{Chem.\ Phys.\ Chem} \textbf{\bibinfo{volume}{14}},
  \bibinfo{pages}{1109} (\bibinfo{year}{2013}).

\bibitem[{\citenamefont{David and Christensen}(2017)}]{DC17}
\bibinfo{author}{\bibfnamefont{C.}~\bibnamefont{David}} \bibnamefont{and}
  \bibinfo{author}{\bibfnamefont{J.}~\bibnamefont{Christensen}},
  \bibinfo{journal}{Appl.\ Phys.\ Lett.} \textbf{\bibinfo{volume}{110}},
  \bibinfo{pages}{261110} (\bibinfo{year}{2017}).

\bibitem[{\citenamefont{Jaklevic and Lambe}(1975)}]{JL1975}
\bibinfo{author}{\bibfnamefont{R.~C.} \bibnamefont{Jaklevic}} \bibnamefont{and}
  \bibinfo{author}{\bibfnamefont{J.}~\bibnamefont{Lambe}},
  \bibinfo{journal}{Phys.\ Rev.\ B} \textbf{\bibinfo{volume}{12}},
  \bibinfo{pages}{4146} (\bibinfo{year}{1975}).

\bibitem[{\citenamefont{H{\"o}vel et~al.}(2010)\citenamefont{H{\"o}vel, Gompf,
  and Dressel}}]{HGD10}
\bibinfo{author}{\bibfnamefont{M.}~\bibnamefont{H{\"o}vel}},
  \bibinfo{author}{\bibfnamefont{B.}~\bibnamefont{Gompf}}, \bibnamefont{and}
  \bibinfo{author}{\bibfnamefont{M.}~\bibnamefont{Dressel}},
  \bibinfo{journal}{Phys.\ Rev.\ B} \textbf{\bibinfo{volume}{81}},
  \bibinfo{pages}{035402} (\bibinfo{year}{2010}).

\bibitem[{\citenamefont{Qian et~al.}(2015)\citenamefont{Qian, Xiao, Lepage,
  Chen, and Liu}}]{QXL15}
\bibinfo{author}{\bibfnamefont{H.}~\bibnamefont{Qian}},
  \bibinfo{author}{\bibfnamefont{Y.}~\bibnamefont{Xiao}},
  \bibinfo{author}{\bibfnamefont{D.}~\bibnamefont{Lepage}},
  \bibinfo{author}{\bibfnamefont{L.}~\bibnamefont{Chen}}, \bibnamefont{and}
  \bibinfo{author}{\bibfnamefont{Z.}~\bibnamefont{Liu}},
  \bibinfo{journal}{Nanophotonics} \textbf{\bibinfo{volume}{4}},
  \bibinfo{pages}{413} (\bibinfo{year}{2015}).

\bibitem[{\citenamefont{Raza et~al.}(2013{\natexlab{b}})\citenamefont{Raza,
  Christensen, Wubs, Bozhevolnyi, and Mortensen}}]{RTW15}
\bibinfo{author}{\bibfnamefont{S.}~\bibnamefont{Raza}},
  \bibinfo{author}{\bibfnamefont{T.}~\bibnamefont{Christensen}},
  \bibinfo{author}{\bibfnamefont{M.}~\bibnamefont{Wubs}},
  \bibinfo{author}{\bibfnamefont{S.~I.} \bibnamefont{Bozhevolnyi}},
  \bibnamefont{and} \bibinfo{author}{\bibfnamefont{N.~A.}
  \bibnamefont{Mortensen}}, \bibinfo{journal}{Phys.\ Rev.\ B}
  \textbf{\bibinfo{volume}{88}}, \bibinfo{pages}{115401}
  (\bibinfo{year}{2013}{\natexlab{b}}).

\bibitem[{\citenamefont{Bondarev and Shalaev}(2017)}]{BS17}
\bibinfo{author}{\bibfnamefont{I.~V.} \bibnamefont{Bondarev}} \bibnamefont{and}
  \bibinfo{author}{\bibfnamefont{V.~M.} \bibnamefont{Shalaev}},
  \bibinfo{journal}{Opt.\ Mater.\ Express} \textbf{\bibinfo{volume}{7}},
  \bibinfo{pages}{3731} (\bibinfo{year}{2017}).

\bibitem[{\citenamefont{Runge and Gross}(1984)}]{RG1984}
\bibinfo{author}{\bibfnamefont{E.}~\bibnamefont{Runge}} \bibnamefont{and}
  \bibinfo{author}{\bibfnamefont{E.~K.~U.} \bibnamefont{Gross}},
  \bibinfo{journal}{Phys.\ Rev.\ Lett.} \textbf{\bibinfo{volume}{52}},
  \bibinfo{pages}{997} (\bibinfo{year}{1984}).

\bibitem[{\citenamefont{Pitarke et~al.}(2007)\citenamefont{Pitarke, Silkin,
  Chulkov, and Echenique}}]{PSC07}
\bibinfo{author}{\bibfnamefont{J.~M.} \bibnamefont{Pitarke}},
  \bibinfo{author}{\bibfnamefont{V.~M.} \bibnamefont{Silkin}},
  \bibinfo{author}{\bibfnamefont{E.~V.} \bibnamefont{Chulkov}},
  \bibnamefont{and} \bibinfo{author}{\bibfnamefont{P.~M.}
  \bibnamefont{Echenique}}, \bibinfo{journal}{Rep.\ Prog.\ Phys.}
  \textbf{\bibinfo{volume}{70}}, \bibinfo{pages}{1} (\bibinfo{year}{2007}).

\bibitem[{\citenamefont{Yan et~al.}(2011)\citenamefont{Yan, Jacobsen, and
  Thygesen}}]{YJT11}
\bibinfo{author}{\bibfnamefont{J.}~\bibnamefont{Yan}},
  \bibinfo{author}{\bibfnamefont{K.~W.} \bibnamefont{Jacobsen}},
  \bibnamefont{and} \bibinfo{author}{\bibfnamefont{K.~S.}
  \bibnamefont{Thygesen}}, \bibinfo{journal}{Phys.\ Rev.\ B}
  \textbf{\bibinfo{volume}{84}}, \bibinfo{pages}{235430}
  (\bibinfo{year}{2011}).

\bibitem[{\citenamefont{Laref et~al.}(2013)\citenamefont{Laref, Cao,
  Asaduzzaman, Runge, Deymier, Ziolkowski, Miyawaki, , and
  Muralidharan}}]{LCA13}
\bibinfo{author}{\bibfnamefont{S.}~\bibnamefont{Laref}},
  \bibinfo{author}{\bibfnamefont{J.}~\bibnamefont{Cao}},
  \bibinfo{author}{\bibfnamefont{A.}~\bibnamefont{Asaduzzaman}},
  \bibinfo{author}{\bibfnamefont{K.}~\bibnamefont{Runge}},
  \bibinfo{author}{\bibfnamefont{P.}~\bibnamefont{Deymier}},
  \bibinfo{author}{\bibfnamefont{R.~W.} \bibnamefont{Ziolkowski}},
  \bibinfo{author}{\bibfnamefont{M.}~\bibnamefont{Miyawaki}}, ,
  \bibnamefont{and}
  \bibinfo{author}{\bibfnamefont{K.}~\bibnamefont{Muralidharan}},
  \bibinfo{journal}{Opt.\ Express} \textbf{\bibinfo{volume}{21}},
  \bibinfo{pages}{11827} (\bibinfo{year}{2013}).

\bibitem[{\citenamefont{Schiller et~al.}(2014)\citenamefont{Schiller,
  El-Fattah, Schirone, Lobo-Checa, Urdanpilleta, Ruiz-Os{\'e}s, Cord{\'o}n,
  Corso, S{\'a}nchez-Portal, Mugarza et~al.}}]{SAS15}
\bibinfo{author}{\bibfnamefont{F.}~\bibnamefont{Schiller}},
  \bibinfo{author}{\bibfnamefont{Z.~M.~A.} \bibnamefont{El-Fattah}},
  \bibinfo{author}{\bibfnamefont{S.}~\bibnamefont{Schirone}},
  \bibinfo{author}{\bibfnamefont{J.}~\bibnamefont{Lobo-Checa}},
  \bibinfo{author}{\bibfnamefont{M.}~\bibnamefont{Urdanpilleta}},
  \bibinfo{author}{\bibfnamefont{M.}~\bibnamefont{Ruiz-Os{\'e}s}},
  \bibinfo{author}{\bibfnamefont{J.}~\bibnamefont{Cord{\'o}n}},
  \bibinfo{author}{\bibfnamefont{M.}~\bibnamefont{Corso}},
  \bibinfo{author}{\bibfnamefont{D.}~\bibnamefont{S{\'a}nchez-Portal}},
  \bibinfo{author}{\bibfnamefont{A.}~\bibnamefont{Mugarza}},
  \bibnamefont{et~al.}, \bibinfo{journal}{New\ J.\ Phys.}
  \textbf{\bibinfo{volume}{16}}, \bibinfo{pages}{123025}
  (\bibinfo{year}{2014}).

\bibitem[{\citenamefont{Sundararaman et~al.}(2018)\citenamefont{Sundararaman,
  Christensen, Ping, Rivera, Joannopoulos, Solja{\v{c}}i{\'c}, and
  Narang}}]{SCP18}
\bibinfo{author}{\bibfnamefont{R.}~\bibnamefont{Sundararaman}},
  \bibinfo{author}{\bibfnamefont{T.}~\bibnamefont{Christensen}},
  \bibinfo{author}{\bibfnamefont{Y.}~\bibnamefont{Ping}},
  \bibinfo{author}{\bibfnamefont{N.}~\bibnamefont{Rivera}},
  \bibinfo{author}{\bibfnamefont{J.~D.} \bibnamefont{Joannopoulos}},
  \bibinfo{author}{\bibfnamefont{M.}~\bibnamefont{Solja{\v{c}}i{\'c}}},
  \bibnamefont{and} \bibinfo{author}{\bibfnamefont{P.}~\bibnamefont{Narang}},
  p. \bibinfo{pages}{arXiv:1806.02672} (\bibinfo{year}{2018}).

\bibitem[{\citenamefont{Shah et~al.}(2018)\citenamefont{Shah, Catellani, Reddy,
  Kinsey, Shalaev, Boltasseva, and Calzolari}}]{SCR18}
\bibinfo{author}{\bibfnamefont{D.}~\bibnamefont{Shah}},
  \bibinfo{author}{\bibfnamefont{A.}~\bibnamefont{Catellani}},
  \bibinfo{author}{\bibfnamefont{H.}~\bibnamefont{Reddy}},
  \bibinfo{author}{\bibfnamefont{N.}~\bibnamefont{Kinsey}},
  \bibinfo{author}{\bibfnamefont{V.}~\bibnamefont{Shalaev}},
  \bibinfo{author}{\bibfnamefont{A.}~\bibnamefont{Boltasseva}},
  \bibnamefont{and}
  \bibinfo{author}{\bibfnamefont{A.}~\bibnamefont{Calzolari}},
  \bibinfo{journal}{ACS\ Photon.} \textbf{\bibinfo{volume}{5}},
  \bibinfo{pages}{2816} (\bibinfo{year}{2018}).

\bibitem[{\citenamefont{Johnson and Christy}(1972)}]{JC1972}
\bibinfo{author}{\bibfnamefont{P.~B.} \bibnamefont{Johnson}} \bibnamefont{and}
  \bibinfo{author}{\bibfnamefont{R.~W.} \bibnamefont{Christy}},
  \bibinfo{journal}{Phys.\ Rev.\ B} \textbf{\bibinfo{volume}{6}},
  \bibinfo{pages}{4370} (\bibinfo{year}{1972}).

\bibitem[{\citenamefont{Jackson}(1999)}]{J99}
\bibinfo{author}{\bibfnamefont{J.~D.} \bibnamefont{Jackson}},
  \emph{\bibinfo{title}{Classical Electrodynamics}}
  (\bibinfo{publisher}{Wiley}, \bibinfo{address}{New York},
  \bibinfo{year}{1999}).

\bibitem[{\citenamefont{{Garc\'{\i}a de Abajo} and
  Manjavacas}(2015)}]{paper254}
\bibinfo{author}{\bibfnamefont{F.~J.} \bibnamefont{{Garc\'{\i}a de Abajo}}}
  \bibnamefont{and}
  \bibinfo{author}{\bibfnamefont{A.}~\bibnamefont{Manjavacas}},
  \bibinfo{journal}{Faraday\ Discuss.} \textbf{\bibinfo{volume}{178}},
  \bibinfo{pages}{87} (\bibinfo{year}{2015}).

\bibitem[{\citenamefont{Profumo et~al.}(2012)\citenamefont{Profumo, Asgari,
  Polini, and MacDonald}}]{PAP12}
\bibinfo{author}{\bibfnamefont{R.~E.~V.} \bibnamefont{Profumo}},
  \bibinfo{author}{\bibfnamefont{R.}~\bibnamefont{Asgari}},
  \bibinfo{author}{\bibfnamefont{M.}~\bibnamefont{Polini}}, \bibnamefont{and}
  \bibinfo{author}{\bibfnamefont{A.~H.} \bibnamefont{MacDonald}},
  \bibinfo{journal}{Phys.\ Rev.\ B} \textbf{\bibinfo{volume}{85}},
  \bibinfo{pages}{085443} (\bibinfo{year}{2012}).

\bibitem[{\citenamefont{Wunsch et~al.}(2006)\citenamefont{Wunsch, Stauber,
  Sols, and Guinea}}]{WSS06}
\bibinfo{author}{\bibfnamefont{B.}~\bibnamefont{Wunsch}},
  \bibinfo{author}{\bibfnamefont{T.}~\bibnamefont{Stauber}},
  \bibinfo{author}{\bibfnamefont{F.}~\bibnamefont{Sols}}, \bibnamefont{and}
  \bibinfo{author}{\bibfnamefont{F.}~\bibnamefont{Guinea}},
  \bibinfo{journal}{New\ J.\ Phys.} \textbf{\bibinfo{volume}{8}},
  \bibinfo{pages}{318} (\bibinfo{year}{2006}).

\bibitem[{\citenamefont{Hwang and {Das Sarma}}(2007)}]{HD07}
\bibinfo{author}{\bibfnamefont{E.~H.} \bibnamefont{Hwang}} \bibnamefont{and}
  \bibinfo{author}{\bibfnamefont{S.}~\bibnamefont{{Das Sarma}}},
  \bibinfo{journal}{Phys.\ Rev.\ B} \textbf{\bibinfo{volume}{75}},
  \bibinfo{pages}{205418} (\bibinfo{year}{2007}).

\bibitem[{\citenamefont{Silveiro et~al.}(2015)\citenamefont{Silveiro, {Plaza
  Ortega}, and {Garc\'{\i}a de Abajo}}}]{paper249}
\bibinfo{author}{\bibfnamefont{I.}~\bibnamefont{Silveiro}},
  \bibinfo{author}{\bibfnamefont{J.~M.} \bibnamefont{{Plaza Ortega}}},
  \bibnamefont{and} \bibinfo{author}{\bibfnamefont{F.~J.}
  \bibnamefont{{Garc\'{\i}a de Abajo}}}, \bibinfo{journal}{Light\ Sci.\ Appl.}
  \textbf{\bibinfo{volume}{4}}, \bibinfo{pages}{e241} (\bibinfo{year}{2015}).

\bibitem[{\citenamefont{Chulkov et~al.}(1999)\citenamefont{Chulkov, Silkin, and
  Echenique}}]{CSE99}
\bibinfo{author}{\bibfnamefont{E.}~\bibnamefont{Chulkov}},
  \bibinfo{author}{\bibfnamefont{V.}~\bibnamefont{Silkin}}, \bibnamefont{and}
  \bibinfo{author}{\bibfnamefont{P.}~\bibnamefont{Echenique}},
  \bibinfo{journal}{Surf.\ Sci.} \textbf{\bibinfo{volume}{437}},
  \bibinfo{pages}{330} (\bibinfo{year}{1999}).

\bibitem[{\citenamefont{Esteban et~al.}(2012)\citenamefont{Esteban, Borisov,
  Nordlander, and Aizpurua}}]{EBN12}
\bibinfo{author}{\bibfnamefont{R.}~\bibnamefont{Esteban}},
  \bibinfo{author}{\bibfnamefont{A.~G.} \bibnamefont{Borisov}},
  \bibinfo{author}{\bibfnamefont{P.}~\bibnamefont{Nordlander}},
  \bibnamefont{and} \bibinfo{author}{\bibfnamefont{J.}~\bibnamefont{Aizpurua}},
  \bibinfo{journal}{Nat.\ Commun.} \textbf{\bibinfo{volume}{3}},
  \bibinfo{pages}{825} (\bibinfo{year}{2012}).

\bibitem[{\citenamefont{Liebsch}(1993)}]{L93}
\bibinfo{author}{\bibfnamefont{A.}~\bibnamefont{Liebsch}},
  \bibinfo{journal}{Phys.\ Rev.\ B} \textbf{\bibinfo{volume}{48}},
  \bibinfo{pages}{11317} (\bibinfo{year}{1993}).

\bibitem[{\citenamefont{Alonso-Gonz{\'a}lez
  et~al.}(2017)\citenamefont{Alonso-Gonz{\'a}lez, Nikitin, Gao, Woessner,
  Lundeberg, Principi, Forcellini, Yan, Sa{\"u}l, Huber et~al.}}]{ANG17}
\bibinfo{author}{\bibfnamefont{P.}~\bibnamefont{Alonso-Gonz{\'a}lez}},
  \bibinfo{author}{\bibfnamefont{A.~Y.} \bibnamefont{Nikitin}},
  \bibinfo{author}{\bibfnamefont{Y.}~\bibnamefont{Gao}},
  \bibinfo{author}{\bibfnamefont{A.}~\bibnamefont{Woessner}},
  \bibinfo{author}{\bibfnamefont{M.~B.} \bibnamefont{Lundeberg}},
  \bibinfo{author}{\bibfnamefont{A.}~\bibnamefont{Principi}},
  \bibinfo{author}{\bibfnamefont{N.}~\bibnamefont{Forcellini}},
  \bibinfo{author}{\bibfnamefont{W.}~\bibnamefont{Yan}},
  \bibinfo{author}{\bibnamefont{Sa{\"u}l}},
  \bibinfo{author}{\bibfnamefont{A.~J.} \bibnamefont{Huber}},
  \bibnamefont{et~al.}, \bibinfo{journal}{Nat.\ Nanotech.}
  \textbf{\bibinfo{volume}{12}}, \bibinfo{pages}{31} (\bibinfo{year}{2017}).

\bibitem[{\citenamefont{Ritchie and Marusak}(1966)}]{RM1966}
\bibinfo{author}{\bibfnamefont{R.~H.} \bibnamefont{Ritchie}} \bibnamefont{and}
  \bibinfo{author}{\bibfnamefont{A.~L.} \bibnamefont{Marusak}},
  \bibinfo{journal}{Surf.\ Sci.} \textbf{\bibinfo{volume}{4}},
  \bibinfo{pages}{234} (\bibinfo{year}{1966}).

\bibitem[{\citenamefont{Ford and Weber}(1984)}]{FW1984}
\bibinfo{author}{\bibfnamefont{G.~W.} \bibnamefont{Ford}} \bibnamefont{and}
  \bibinfo{author}{\bibfnamefont{W.~H.} \bibnamefont{Weber}},
  \bibinfo{journal}{Phys.\ Rep.} \textbf{\bibinfo{volume}{113}},
  \bibinfo{pages}{195} (\bibinfo{year}{1984}).

\bibitem[{\citenamefont{{Garc\'{\i}a de Abajo}}(2008)}]{paper119}
\bibinfo{author}{\bibfnamefont{F.~J.} \bibnamefont{{Garc\'{\i}a de Abajo}}},
  \bibinfo{journal}{J.\ Phys.\ Chem.\ C} \textbf{\bibinfo{volume}{112}},
  \bibinfo{pages}{17983} (\bibinfo{year}{2008}).

\bibitem[{\citenamefont{Lindhard}(1954)}]{L1954}
\bibinfo{author}{\bibfnamefont{J.}~\bibnamefont{Lindhard}},
  \bibinfo{journal}{K.\ Dan.\ Vidensk.\ Selsk.\ Mat.\ Fys.\ Medd.}
  \textbf{\bibinfo{volume}{28}}, \bibinfo{pages}{1} (\bibinfo{year}{1954}).

\bibitem[{\citenamefont{Hedin and Lundqvist}(1970)}]{HL1970}
\bibinfo{author}{\bibfnamefont{L.}~\bibnamefont{Hedin}} \bibnamefont{and}
  \bibinfo{author}{\bibfnamefont{S.}~\bibnamefont{Lundqvist}}, in
  \emph{\bibinfo{booktitle}{Solid State Physics}}, edited by
  \bibinfo{editor}{\bibfnamefont{D.~T.} \bibnamefont{Frederick~Seitz}}
  \bibnamefont{and}
  \bibinfo{editor}{\bibfnamefont{H.}~\bibnamefont{Ehrenreich}}
  (\bibinfo{publisher}{Academic Press}, \bibinfo{year}{1970}),
  vol.~\bibinfo{volume}{23} of \emph{\bibinfo{series}{Solid State Physics}},
  pp. \bibinfo{pages}{1 -- 181}.

\bibitem[{\citenamefont{Mermin}(1970)}]{M1970}
\bibinfo{author}{\bibfnamefont{N.~D.} \bibnamefont{Mermin}},
  \bibinfo{journal}{Phys.\ Rev.\ B} \textbf{\bibinfo{volume}{1}},
  \bibinfo{pages}{2362} (\bibinfo{year}{1970}).

\bibitem[{\citenamefont{McMahon et~al.}(2010)\citenamefont{McMahon, Gray, and
  Schatz}}]{MGS10_2}
\bibinfo{author}{\bibfnamefont{J.~M.} \bibnamefont{McMahon}},
  \bibinfo{author}{\bibfnamefont{S.~K.} \bibnamefont{Gray}}, \bibnamefont{and}
  \bibinfo{author}{\bibfnamefont{G.~C.} \bibnamefont{Schatz}},
  \bibinfo{journal}{Nano\ Lett.} \textbf{\bibinfo{volume}{10}},
  \bibinfo{pages}{3473} (\bibinfo{year}{2010}).

\bibitem[{\citenamefont{Lin et~al.}(2017)\citenamefont{Lin, Fan, and
  Liu}}]{LFL17}
\bibinfo{author}{\bibfnamefont{I.-T.} \bibnamefont{Lin}},
  \bibinfo{author}{\bibfnamefont{C.}~\bibnamefont{Fan}}, \bibnamefont{and}
  \bibinfo{author}{\bibfnamefont{J.-M.} \bibnamefont{Liu}},
  \bibinfo{journal}{IEEE\ J.\ Sel.\ Top.\ Quant.\ Electr.}
  \textbf{\bibinfo{volume}{23}}, \bibinfo{pages}{144} (\bibinfo{year}{2017}).

\bibitem[{\citenamefont{Pines and Nozi\`{e}res}(1966)}]{PN1966}
\bibinfo{author}{\bibfnamefont{D.}~\bibnamefont{Pines}} \bibnamefont{and}
  \bibinfo{author}{\bibfnamefont{P.}~\bibnamefont{Nozi\`{e}res}},
  \emph{\bibinfo{title}{The Theory of quantum liquids}} (\bibinfo{publisher}{W.
  A. Benjamin, Inc.}, \bibinfo{address}{New York}, \bibinfo{year}{1966}).

\bibitem[{\citenamefont{de~Vega and {Garc\'{\i}a de Abajo}}(2017)}]{paper295}
\bibinfo{author}{\bibfnamefont{S.}~\bibnamefont{de~Vega}} \bibnamefont{and}
  \bibinfo{author}{\bibfnamefont{F.~J.} \bibnamefont{{Garc\'{\i}a de Abajo}}},
  \bibinfo{journal}{ACS\ Photon.} \textbf{\bibinfo{volume}{4}},
  \bibinfo{pages}{2367} (\bibinfo{year}{2017}).

\bibitem[{\citenamefont{Paniago et~al.}(1995)\citenamefont{Paniago, Matzdorf,
  Meister, and Goldmann}}]{PMM95}
\bibinfo{author}{\bibfnamefont{R.}~\bibnamefont{Paniago}},
  \bibinfo{author}{\bibfnamefont{R.}~\bibnamefont{Matzdorf}},
  \bibinfo{author}{\bibfnamefont{G.}~\bibnamefont{Meister}}, \bibnamefont{and}
  \bibinfo{author}{\bibfnamefont{A.}~\bibnamefont{Goldmann}},
  \bibinfo{journal}{Surf.\ Sci.} \textbf{\bibinfo{volume}{336}},
  \bibinfo{pages}{113} (\bibinfo{year}{1995}).

\bibitem[{\citenamefont{{Garc\'{\i}a de Abajo}}(2010)}]{paper149}
\bibinfo{author}{\bibfnamefont{F.~J.} \bibnamefont{{Garc\'{\i}a de Abajo}}},
  \bibinfo{journal}{Rev.\ Mod.\ Phys.} \textbf{\bibinfo{volume}{82}},
  \bibinfo{pages}{209} (\bibinfo{year}{2010}).

\bibitem[{\citenamefont{Hwang and Sarma}(2007)}]{HS07}
\bibinfo{author}{\bibfnamefont{E.~H.} \bibnamefont{Hwang}} \bibnamefont{and}
  \bibinfo{author}{\bibfnamefont{S.~D.} \bibnamefont{Sarma}},
  \bibinfo{journal}{Phys.\ Rev.\ B} \textbf{\bibinfo{volume}{75}},
  \bibinfo{pages}{205418} (\bibinfo{year}{2007}).

\bibitem[{\citenamefont{Clementi and Roetti}(1974)}]{CR1974}
\bibinfo{author}{\bibfnamefont{E.}~\bibnamefont{Clementi}} \bibnamefont{and}
  \bibinfo{author}{\bibfnamefont{C.}~\bibnamefont{Roetti}},
  \bibinfo{journal}{At.\ Data\ Nucl.\ Data\ Tables}
  \textbf{\bibinfo{volume}{14}}, \bibinfo{pages}{177} (\bibinfo{year}{1974}).

\bibitem[{\citenamefont{Geick et~al.}(1966)\citenamefont{Geick, Perry, and
  Rupprecht}}]{GPR1966}
\bibinfo{author}{\bibfnamefont{R.}~\bibnamefont{Geick}},
  \bibinfo{author}{\bibfnamefont{C.~H.} \bibnamefont{Perry}}, \bibnamefont{and}
  \bibinfo{author}{\bibfnamefont{G.}~\bibnamefont{Rupprecht}},
  \bibinfo{journal}{Phys.\ Rev.} \textbf{\bibinfo{volume}{146}},
  \bibinfo{pages}{543} (\bibinfo{year}{1966}).

\bibitem[{\citenamefont{Golla et~al.}(2013)\citenamefont{Golla, Chattrakun,
  Watanabe, Taniguchi, LeRoy, and Sandhu}}]{GCW13}
\bibinfo{author}{\bibfnamefont{D.}~\bibnamefont{Golla}},
  \bibinfo{author}{\bibfnamefont{K.}~\bibnamefont{Chattrakun}},
  \bibinfo{author}{\bibfnamefont{K.}~\bibnamefont{Watanabe}},
  \bibinfo{author}{\bibfnamefont{T.}~\bibnamefont{Taniguchi}},
  \bibinfo{author}{\bibfnamefont{B.~J.} \bibnamefont{LeRoy}}, \bibnamefont{and}
  \bibinfo{author}{\bibfnamefont{A.}~\bibnamefont{Sandhu}},
  \bibinfo{journal}{Appl.\ Phys.\ Lett.} \textbf{\bibinfo{volume}{102}},
  \bibinfo{pages}{161906} (\bibinfo{year}{2013}).

\bibitem[{\citenamefont{Kevan and Gaylord}(1987)}]{KG1987}
\bibinfo{author}{\bibfnamefont{S.~D.} \bibnamefont{Kevan}} \bibnamefont{and}
  \bibinfo{author}{\bibfnamefont{R.~H.} \bibnamefont{Gaylord}},
  \bibinfo{journal}{Phys.\ Rev.\ B} \textbf{\bibinfo{volume}{36}},
  \bibinfo{pages}{5809} (\bibinfo{year}{1987}).

\bibitem[{\citenamefont{Zeman and Schatz}(1987)}]{ZS1987}
\bibinfo{author}{\bibfnamefont{E.~J.} \bibnamefont{Zeman}} \bibnamefont{and}
  \bibinfo{author}{\bibfnamefont{G.~C.} \bibnamefont{Schatz}},
  \bibinfo{journal}{J.\ Phys.\ Chem.} \textbf{\bibinfo{volume}{91}},
  \bibinfo{pages}{634} (\bibinfo{year}{1987}).

\end{thebibliography}

\end{document}